\documentclass[aps,10pt,a4paper,prb,superscriptaddress,twocolumn,showkeys,longbibliography,floatfix]%
{revtex4-1}

\usepackage{xcolor}

\usepackage{dcolumn}

\usepackage[normalem]{ulem}

\usepackage{bm}
\usepackage{amssymb}
\usepackage{amsmath}

\usepackage[english]{babel}
\usepackage{graphicx} %[draft] place holder for graphs
\usepackage[breaklinks=true, colorlinks=true, linkcolor=blue, urlcolor=blue, citecolor=blue]{hyperref}
\newcommand\units[1]{\ensuremath{\,\mathrm{#1}}}
\DeclareMathOperator{\sgn}{sgn}

\providecommand{\f}[2]{#1\left(#2\right)}
\providecommand{\ls}[2]{#1_\mathrm{#2}}

\definecolor{pinegreen}{rgb}{0.0, 0.47, 0.44}
\definecolor{persiangreen}{rgb}{0.0, 0.65, 0.58}
\definecolor{pakistangreen}{rgb}{0.0, 0.4, 0.0}
\definecolor{mossgreen}{rgb}{0.68, 0.87, 0.68}
\definecolor{olivegreen}{rgb}{0.58, 0.57, 0.27}
\definecolor{yelloworange}{rgb}{1.0, 0.68, 0.26}
\definecolor{purple}{rgb}{0.63, 0.13, 0.94}
\definecolor{brickred}{rgb}{0.8, 0.25, 0.33}
\definecolor{palecarmine}{rgb}{0.69, 0.25, 0.21}
\definecolor{bondiblue}{rgb}{0.0, 0.58, 0.71}
\definecolor{awesome}{rgb}{1.0, 0.13, 0.32}
\definecolor{carnationpink}{rgb}{1.0, 0.65, 0.79}
\definecolor{airforceblue}{rgb}{0.36, 0.54, 0.66}
\definecolor{beaublue}{rgb}{0.74, 0.83, 0.9}
\definecolor{bleudefrance}{rgb}{0.19, 0.55, 0.91}
\definecolor{blue(pigment)}{rgb}{0.2, 0.2, 06}
\definecolor{brightlavender}{rgb}{0.75, 0.58, 0.89}
\definecolor{brightgreen}{rgb}{0.4, 1.0, 0.0}
\definecolor{caribbeangreen}{rgb}{0.0, 0.8, 0.6}

\RequirePackage{xspace}

\newif\ifcom
\newif\ifdel
\comtrue
\deltrue

\begin{document}
\title{Temporal evolution of electric transport properties of YBCO Josephson junctions produced by focused Helium ion beam irradiation}

\author{M.~Karrer}
%\email{max.karrer@uni-tuebingen.de}
%\thanks{}
\affiliation{%
  Physikalisches Institut, Center for Quantum Science (CQ) and LISA$^+$,
  Eberhard Karls Universit\"at T\"ubingen,
  Auf der Morgenstelle 14,
  72076 T\"ubingen, Germany
}

\author{K.~Wurster}
%\email{katja.wurster@uni-tuebingen.de}
\affiliation{%
  Physikalisches Institut, Center for Quantum Science (CQ) and LISA$^+$,
  Eberhard Karls Universit\"at T\"ubingen,
  Auf der Morgenstelle 14,
  72076 T\"ubingen, Germany
}

\author{J.~Linek}
%\email{julian.linek@uni-tuebingen.de}
\affiliation{%
  Physikalisches Institut, Center for Quantum Science (CQ) and LISA$^+$,
  Eberhard Karls Universit\"at T\"ubingen,
  Auf der Morgenstelle 14,
  72076 T\"ubingen, Germany
}

\author{M.~Meichsner}
%\email{moritz.meichsner@student.uni-tuebingen.de}
\affiliation{%
  Physikalisches Institut, Center for Quantum Science (CQ) and LISA$^+$,
  Eberhard Karls Universit\"at T\"ubingen,
  Auf der Morgenstelle 14,
  72076 T\"ubingen, Germany
}

\author{R.~Kleiner}
%\email{kleiner@uni-tuebingen.de}
\affiliation{%
  Physikalisches Institut, Center for Quantum Science (CQ) and LISA$^+$,
  Eberhard Karls Universit\"at T\"ubingen,
  Auf der Morgenstelle 14,
  72076 T\"ubingen, Germany
}

\author{E.~Goldobin}
%\email{gold@uni-tuebingen.de}
\affiliation{%
  Physikalisches Institut, Center for Quantum Science (CQ) and LISA$^+$,
  Eberhard Karls Universit\"at T\"ubingen,
  Auf der Morgenstelle 14,
  72076 T\"ubingen, Germany
}

\author{D.~Koelle}
\email{koelle@uni-tuebingen.de}
\affiliation{%
  Physikalisches Institut, Center for Quantum Science (CQ) and LISA$^+$,
  Eberhard Karls Universit\"at T\"ubingen,
  Auf der Morgenstelle 14,
  72076 T\"ubingen, Germany
}

\date{\today}

%%%%%%%%%%%%%%%%%%%%%%%%%%%%%%
\begin{abstract}

Using a $30\units{keV}$ focused He ion beam (He-FIB) with a wide range of irradiation doses $D=100$ to $1000\units{ions/nm}$ we fabricated Josephson and resistive barriers within microbridges of epitaxially grown single crystalline {YBa$_2$Cu$_3$O$_{7-\delta}$} (YBCO) thin films and investigated the change of their electric transport properties with time.
One set of samples (\#1A) was simply stored at room temperature under nitrogen atmosphere.
A second set (\#2D) was post-annealed at $90^\circ\units{C}$ using high oxygen pressures and a third set (\#2E) at low oxygen pressures.
We found that for \#1A the critical current density $\ls{j}{c}$ at $4.2\units{K}$ changes as $\ls{j}{c}\propto\exp(-\sqrt{t/\tau})$ with time $t$, where the relaxation times $\tau$ increases exponentially with $D$, which can be described within a diffusion based model.
In order to increase the diffusion rate we annealed the junctions from \#2D at $90^\circ\units{C}$ for $30\units{min}$ in oxygen environment.
Directly after annealing the critical current density $\ls{j}{c}$ increased, while the normal state resistance $\ls{R}{n}$ decreased.
Repeated measurements showed that within a week the junctions relaxed to a quasi-stable state, in which the time scale for junction parameter variations increased to several weeks, making this a feasible option to achieve temporal stability of parameters of He-FIB Josephson junctions in YBCO.
\end{abstract}
%%%%%%%%%%%%%%%%%%%%%%%%%%%%%%

%\keywords{Josephson junction, Annealing, focused ion beam}
%Use showkeys class option if keyword display desired

\maketitle

%%%%%%%%%%%%%%%%%%%%%%%%%%%%%%%%%
\section{Introduction}
\label{sec:introduction}
%%%%%%%%%%%%%%%%%%%%%%%%%%%%%%%%%

The manufacturing of superconducting electronics devices and circuits based on the high-transition temperature (high-$\ls{T}{c}$) cuprate superconductors is still challenging.
Using a multilayer technology including trilayer Josephson junctions (JJs), difficulties arise due to complex material composition and the need for fully epitaxial layer growth \cite{Koelle99,Ludwig95}.
So far, high-$\ls{T}{c}$ JJ technologies are vastly based on single layer YBCO thin films with grain boundaries as Josephson barriers \cite{Hilgenkamp02, Gross97}.
However, to rely on grain boundaries is heavily restricting with respect to possible circuit designs.
In the 1990s, another approach was pursued with various attempts to create Josephson barriers in epitaxially grown YBCO films by local irrradiation with high-energetic electron or ion beams \cite{Bergeal2007,Kang2002,Tolpygo93,Booij1997}.
However, the resulting barriers were lacking quality with respect to lower values of the characteristic voltage $V_{\rm c}$ (critical current times junction resistance) and greater excess currents compared to grain-boundary-based JJs.
After the invention of the gas field ion source (GFIS), which enabled unprecedentedly high spatial resolution in FIB imaging and patterning with $30\units{keV}$ helium ions, a breakthrough with respect to JJs occured in 2015 \cite{Cybart15}.
Since then, the fabrication of JJs and more sophisticated devices by irradiating high-$T_c$ cuprate superconductors with a $30\units{keV}$ He-FIB was reported by several groups \cite{Cho18, Mueller19, Couedo20, Chen22, Merino23}.

The mechanism behind the fabrication of He-FIB-induced Josephson barriers in YBCO is that, presumably, the accelerated He ions with an energy of $30\units{keV}$ and thus their significant larger momentum compared to electrons, are capable of displacing the atoms in the YBCO lattice.
SRIM \cite{Ziegler10} simulations show that mainly the chain oxygen, with $\sim 1\units{eV}$ binding energy \cite{Cui92} is displaced, which creates oxygen vacancies in YBCO \cite{Lang12}.
Since superconductivity in YBCO heavily depends on the oxygen doping, one can expect a local reduction of the superconducting properties \cite{Mletschnig19,Aichner19,Aichner20}.
He-FIB irradiation will therefore introduce a local imbalance, and with time this imbalance will be reduced due to thermally activated diffusion, which will drive the system back towards a state of (quasi-)equilibrium.

Obviously, the temporal evolution of the He-FIB-induced defect structure at the barrier will directly affect the electronic characteristics of the devices, i.e., device properties changes with time.
This effect is certainly detrimental to applications where stable device properties are required.
It has been shown already that devices can be stabilized by annealing after irradiation.
So far, annealing of devices in YBCO was performed for devices produced by irradiation with $(80\text{--}120)\units{keV}$ electrons with a scanning transmission electron microscope \cite{Tolpygo93}
and $175\units{keV}$ Ne+ ions using a trilayer implantation mask \cite{Cho16}.
However, the long-term temporal stability of parameters describing the electrical transport properties of YBCO JJs produced by He-FIB irradiation has not yet been investigated.
Temporal stability is important not only for applications where stable device parameters are needed over time but also for research, e.g.~in concurrent measurements that need some time to setup.
An easy known way for achieving temporal stability of He-FIB JJs is to store them at cryogenic temperatures.
However, this method is under most circumstances not very suitable.

In this paper, we present the results obtained with a different approach:
We first investigated the temporal evolution of electric transport parameters of He-FIB JJs which were stored at room temperature in a N$_2$ atmosphere.
Subsequently, we investigated how an annealing process with different O$_2$ partial pressures changes the junction properties and their temporal evolution.

%%%%%%%%%%%%%%%%%%%%%%%%%%%%%%%%%
\section{Sample Fabrication}
\label{sec:fabrication}
%%%%%%%%%%%%%%%%%%%%%%%%%%%%%%%%%

We fabricated epitaxially grown $c$-axis oriented YBCO thin films on $10\times 10\units{mm^2}$ single-crystal (100) (LaAlO3)$_{0.3}$(Sr2AlTaO6)$_{0.7}$ (LSAT) substrates by pulsed laser deposition with a 20-nm-thick in-situ Au layer on top deposited by electron beam evaporation.
For this study we used two chips (\#1 and \#2), which where fabricated with nominally the same process parameters.
To estimate the quality of the unpatterned YBCO films, x-ray diffraction (XRD) was performed after film growth, and the critical temperature $\ls{T}{c}$ was measured using an inductive method.

%%%%%%%%%%% Fig.1 %%%%%%%%%%%%%%%%%%%%%%%%%%%%%%%%%%%%%
\begin{figure}[!htb]
\begin{center}
\includegraphics*[width=1\columnwidth]{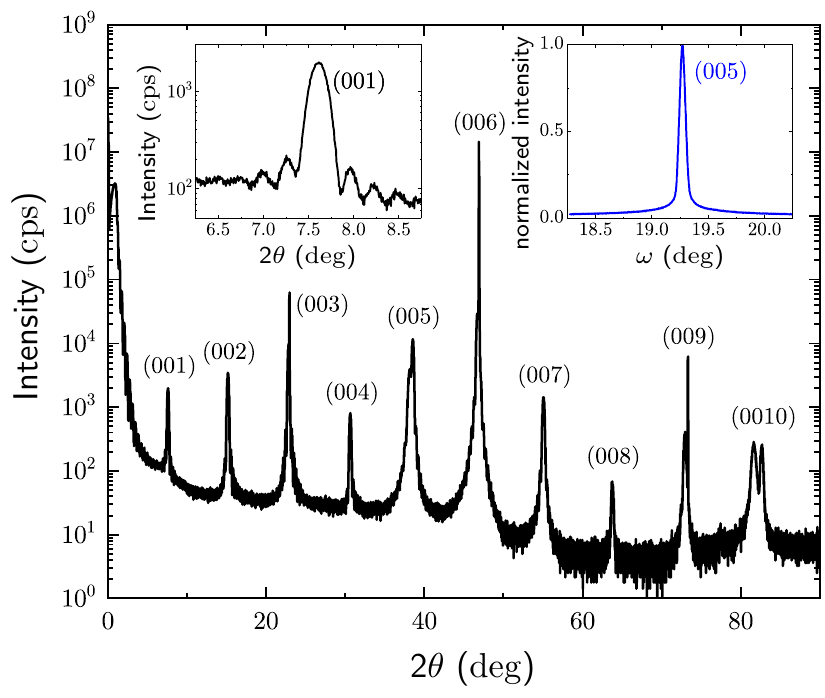}%{XRD}
\end{center}
\caption{
XRD data of the unpatterned chip \#2.
Main graph: $2\theta$-$\omega$ overview scan with $(00\ell)$ Bragg peaks.
Left inset: zoom of the YBCO (001) peak with Laue fringes.
Right inset: $\omega$ scan (rocking curve) at the YBCO (005) peak.
}
\label{fig:xrd}
\end{figure}
%%%%%%%%%%% Fig.1 %%%%%%%%%%%%%%%%%%%%%%%%%%%%%%%%%%%%%

Figure \ref{fig:xrd} shows a representative set of XRD measurements from chip \#2.
From the positions of the ($00\ell$) Bragg peaks along the $2\theta$-axis, the $c$-axis lattice constant of the YBCO film was determined to be $11.67\units{\AA}$.
A YBCO film thickness $d=30\units{nm}$ could be extracted from the Laue fringes \cite{Miller22} at the YBCO (001) Bragg peak (see left inset in \ref{fig:xrd}).
The right inset in \ref{fig:xrd} shows the rocking curve of the YBCO (005) Bragg peak.
The full width at half maximum (FWHM) of the rocking curve is $0.069^\circ$, thus indicating a high crystalline quality.
Our films yield $\ls{T}{c}\approx 89\units{K}$, with a transition width below $1\units{K}$.
Both YBCO films showed variations of their properties ($\ls{T}{c}$, thickness, FWHM of the rocking curve and $c$-axis lattice constant) below 5\% of the given values.

%%%%%%%%%%% Fig.2 %%%%%%%%%%%%%%%%%%%%%%%%%%%%%%%%%%%%%
\begin{figure}[!htb]
  \begin{center}
    \includegraphics*[width=1\columnwidth]{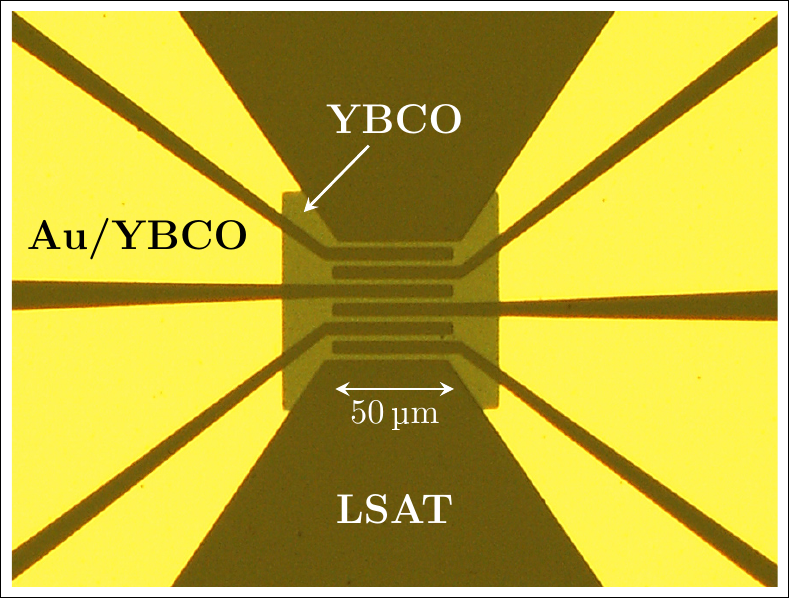}%{layout}
  \end{center}
  \caption{
    Optical microscope image of the central part of structure \#2A with seven microbridges after prepatterning.
  }
  \label{fig:layout}
\end{figure}
%%%%%%%%%%% Fig.2 %%%%%%%%%%%%%%%%%%%%%%%%%%%%%%%%%%%%%

For a two-step photolithography process we used a \textsc{MLA100 maskless aligner} (Heidelberg Instruments) and positive resist \textsc{ma-P 1205} (Micro Resist Technology).
During the first step we patterned the Au/YBCO bilayer by Ar ion milling to create $w=3\units{\mu m}$ wide bridges and contact pads covered by Au.
During the second step we removed the Au layer with commercially available Lugol's iodine \textsc{TechniEtch\textsuperscript{TM}ACI2} only on top of the micro bridges to expose the YBCO underneath, and along narrow stripes across the contact pads to split the Au contacts into two parts, enabling 4-point measurements of the junctions later on.
On each chip, we fabricated $8$  structures (A to H), each of which contain seven microbridges (each $50\units{\mu m}$ long) in a meandering configuration.
Fig.~\ref{fig:layout} shows an optical microscope image after lithography of the central part of chip \#2, structure A, denoted as \#2A.

%%%%%%%%%%% Tab.1 %%%%%%%%%%%%%%%%%%%%%%%%%%%%%%%%%%%%%
\begin{table}
  \caption{Dose values used to generate JJs in \#1A for storage at room temperature  and \#2D, \#2E for annealing.}
  \label{tab:dose-values}
  \begin{tabular}{lc}
    \hline
    \textbf{Junction} & \textbf{Dose}\\
    \#1A & \units{ions/nm}\\
    \hline
    -1 & 100\\
    -2 & 200\\
    -3 & 250\\
    -4 & 400\\
    -5 & 600\\
    -6 & 800\\
    -7 & 1000\\
    \hline
  \end{tabular}
  \qquad
  \begin{tabular}{lc}
    \hline
    \textbf{Junction} & \textbf{Dose}\\
    \#2D, \#2E & \units{ions/nm}\\
    \hline
    -1 & 200\\
    -2 & 400\\
    -3 & 500\\
    -4 & 600\\
    -5 & 700\\
    -6 & 800\\
    -7 & 1000\\
    \hline
  \end{tabular}
\end{table}
%%%%%%%%%%% Tab.1 %%%%%%%%%%%%%%%%%%%%%%%%%%%%%%%%%%%%%

For He-FIB irradiation, the samples were mounted (and electrically grounded) on an aluminum sample stub and loaded into a \textsc{ORION NanoFab} (Zeiss Microscopy).
Junctions with different line dose $D$ (see Tab.~\ref{tab:dose-values}) were written by irradiating with a $30\units{keV}$ He-FIB along a line perpendicular to the YBCO microbridges.
The dwell time per pixel was set to $1\units{\mu s}$ and the number of line repeats was varied to set different dose values in units of ions per nm.

%%%%%%%%%%%%%%%%%%%%%%%%%%%%%%%%%
\section{Experimental Setup and procedure}
\label{sec:setup}
%%%%%%%%%%%%%%%%%%%%%%%%%%%%%%%%%

We measured $I$--$V$ characteristic (IVC) and the modulation of the critical current $\ls{I}{c}$ in an externally applied magnetic field $B$ (along the normal to the YBCO film plane) in a current bias mode.
For the $\f{\ls{I}{c}}{B}$ measurements a voltage criterion of $3\units{\mu V}$ was used to determine $\ls{I}{c}$.
This criterion was $1\units{\mu V}$ above the typical noise level in our setup for the used gain settings of the preamplifiers.
Measurements were performed by immersing the samples in liquid helium at temperature $T=4.2\units{K}$ inside an electrically and magnetically shielded environment within a high frequency shielding chamber.

In order to investigate the evolution of the properties of the junctions over time, we define day zero ($t=0$) as the day when the bridges have been He-FIB-irradiated to form the JJs.
Due to inevitable delays in the handling of the sample, the first measurement after the irradiation is conducted at $0 < \ls{t}{initial} \le 1\units{day}$.
Furthermore, for the compensation of errors on the time axis we tracked (and subtracted) the time when the sample was in the measurement setup and exposed to lower temperatures.

\subsection{Temporal evolution of the properties of junctions stored at room temperature}

Chip \#1 was stored in a flow box under N$_2$ atmosphere between measurements to avoid degradation of the YBCO surface due to air humidity \cite{Hooker93}.
Subsequently, and repeatedly, the JJ properties have been measured.
In the beginning, measurements where performed more frequently for a better resolution of the larger rate of change of JJ properties right after irradiation.

\subsection{Annealing at elevated temperatures}

To stabilize the JJ properties, we annealed samples for a certain time $\ls{t}{a}$ in a oxygen partial pressure $p_{\rm O_2}$.
Annealing was performed in a vacuum (annealing) chamber with inlets for O$_2$ and N$_2$, with the sample placed on a resistive heating element with an integrated temperature sensor.
A digital temperature controller \textsc{OMRON E5AC} was used to set and keep the sensor temperature $\ls{T}{a}$ constant during annealing.
After sample loading, the annealing chamber was evacuated to an absolute pressure $<0.1\units{mbar}$, then flooded with oxygen and evacuated to $<0.1\units{mbar}$ again to flush the chamber and the pipes from the gas bottle to the annealing system.
After this procedure, $p_{\rm O_2}$ was set to the desired value and the temperature was ramped up to $\ls{T}{a}=90^\circ\units{C}$ with a rate of $20^\circ \units{C/min}$.
Pristine YBCO remains unharmed at $90^\circ\units{C}$ with no degradation even in vacuum \cite{Mogro-Campero95}.

After writing barriers into structure \#2D, the chip \#2 was annealed at $p_{\rm O_2}=950\units{mbar}$ for $\ls{t}{a}=30\units{min}$ before the annealing chamber was flooded with O$_2$ to ambient pressure.
Subsequently, the sample was placed onto a stainless steel plate for cooldown, outside the chamber in air.
After completing measurements on structure \#2D we wrote barriers into structure \#2E and annealed chip \#2 in vacuum.
For this we just left the vacuum pump of the annealing chamber connected after the flushing procedure, i.e., during the subsequent annealing process for $\ls{t}{a}=30\units{min}$ we had $p_{\rm O_2} < 0.1\units{mbar}$.
Subsequently, we switched off the heater, vented the chamber with N$_2$ and then placed the sample onto a stainless steel plate for cooldown, outside the chamber in air.

Samples were measured directly after irradiation (at $t = 0$), have been annealed with the corresponding oxygen partial  pressure at  $t = 1\units{day}$ and measured subsequently all within two days.
After this, the chip was stored under N$_2$ environment at room temperature and the junctions were repeatedly measured.

\subsection{Evaluation of IVC}

\begin{figure}[!htb]
\begin{center}
\includegraphics*[width=1\columnwidth]{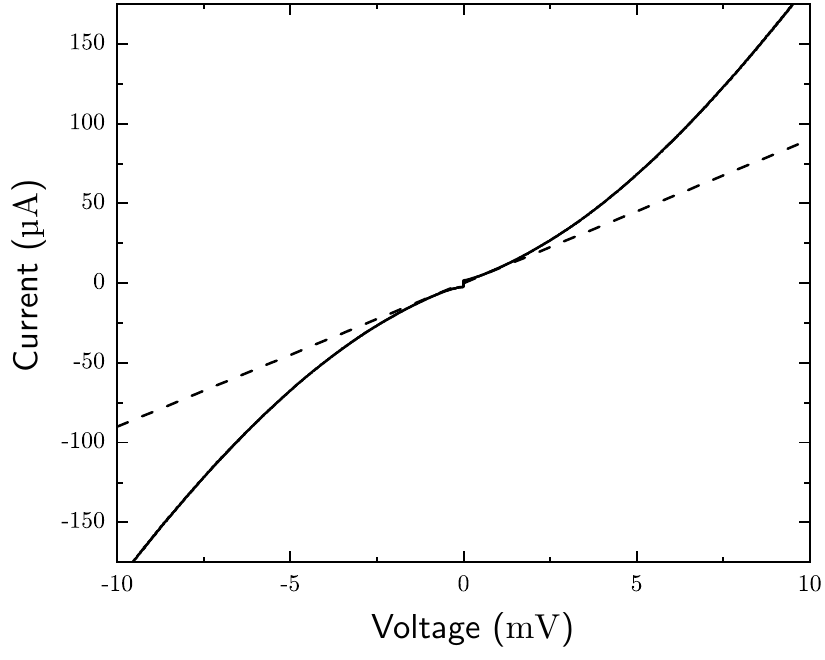}%{IV-10A7}
\end{center}
\caption{
IVC of Junction \#1A-7 taken at the first minimum of $\f{\ls{I}{c}}{B}$ after $t=160\units{days}$ clearly bending away from an ohmic resistance (dashed line) for larger voltages.
}
  \label{fig:IV-10A7}
\end{figure}

For junctions irradiated with $D\gtrsim 800\units{ions/nm}$, %for \#1A and $>X00\units{ions/nm}$ for \#2B,
i.e., close to the transition from medium to high dose \cite{Mueller19}, the IVC exhibit a $I\sim V^3$ contribution, in addition to a linear scaling $I\sim V$ (see Fig.~\ref{fig:IV-10A7}), while for junctions with $D<500\units{ions/nm}$ %for \#1A and $<500\units{ions/nm}$ for \#2B
IVC exhibit an excess current $\ls{I}{ex}$.
%
% Bei A ist die Grenze für I_ex zwischen A4 (I_ex >0) und A5 (I_ex~0).
% Für D ist die Grenze zw. D2 (I_ex >0) und D3 (I_ex~0).
% Für den V^3 Teil muss ich mir erst noch ein Kriterium überlegen.
% Rein visuell weisen A7, D6 und D7 eine V^3 Krümmung auf.
%
In order to consistently determine the normal state resistance $\ls{R}{n}$ for all junctions, we applied an external magnetic field to suppress the critical current  $\ls{I}{c}$ and used the functional dependence
%
%%%%%%%%%%%%%%%%%%%%%%%%%%%%%%%%%%%%%%%%
\begin{equation}\label{Eq:fitting}
  \f{I}{V} = a\cdot V^3 + \frac{V}{\ls{R}{n}}  + \sgn(V)\cdot \ls{I}{ex}
\end{equation}
%%%%%%%%%%%%%%%%%%%%%%%%%%%%%%%%%%%%%%%%
%
to fit all our measured IVC.
Since $\ls{I}{c}$ was never fully suppressed to zero, we discarded parts of the IVC where $\left|V\right| \leq 1\units{mV}$ for the evaluation of $\ls{R}{n}$.
Here, the $1\units{mV}$ criterion was used, as this was the typical voltage level, above which the IVC did coincide for any value of applied external magnetic field.

%%%%%%%%%%% Fig.3 %%%%%%%%%%%%%%%%%%%%%%%%%%%%%%%%%%%%%
%%%%%%%%%%% Fig.3 %%%%%%%%%%%%%%%%%%%%%%%%%%%%%%%%%%%%%

The maximum critical current $\ls{I}{c,max}$ for each junction was determined from the maximum of the recorded $\f{\ls{I}{c}}{B}$ curves.
Note that due to thermal noise rounding, the experimentally determined $\ls{I}{c,max}$ can be significantly suppressed below the noise-free maximum critical current $\ls{I}{0}$.
From numerical simulations of overdamped JJs within the RCSJ model, one knows that for a noise parameter $\Gamma\equiv\ls{I}{th}/\ls{I}{0}=0.05$, with $\ls{I}{th}=2\pi k_\mathrm{b}T/\ls{\Phi}{0}$, $\ls{I}{c,max}$ of a non-hysteretic junction is suppressed by $\sim 10\units{\%}$ ($\ls{\Phi}{0}$ is the magnetic flux quantum and $k_\mathrm{b}$ is the Boltzmann constant) \cite{Chesca-SHB-2}.
With $\ls{I}{th}\approx 176\units{nA}$ at $T=4.2\units{K}$, a noise parameter $\Gamma=0.05$ corresponds to $\ls{I}{0}\approx 3.5\units{\mu A}$, i.e., significant noise rounding is only expected for JJs with critical currents below a few \units{\mu A}, which is only relevant for the junction \#1A7, for which the measured $\ls{I}{c,max}$ is a lower bound.

%%%%%%%%%%%%%%%%%%%%%%%%%%%%%%%%%
\section{Results and discussion}
\label{sec:results}
%%%%%%%%%%%%%%%%%%%%%%%%%%%%%%%%%

Now we discuss and analyze the evolution of the electric transport properties of He-FIB JJs with time.
Note that unless stated otherwise, for simplicity, we denote now the maximum (with respect to variable magnetic field $B$) critical current as $\ls{I}{c}$ and the corresponding maximum critical current density as $\ls{j}{c}=\ls{I}{c}/(wd)$.

\subsection{Temporal evolution of electric transport properties of He-FIB JJs stored at room temperature}

\begin{figure}[!htb]
\begin{center}
\includegraphics*[width=1\columnwidth]{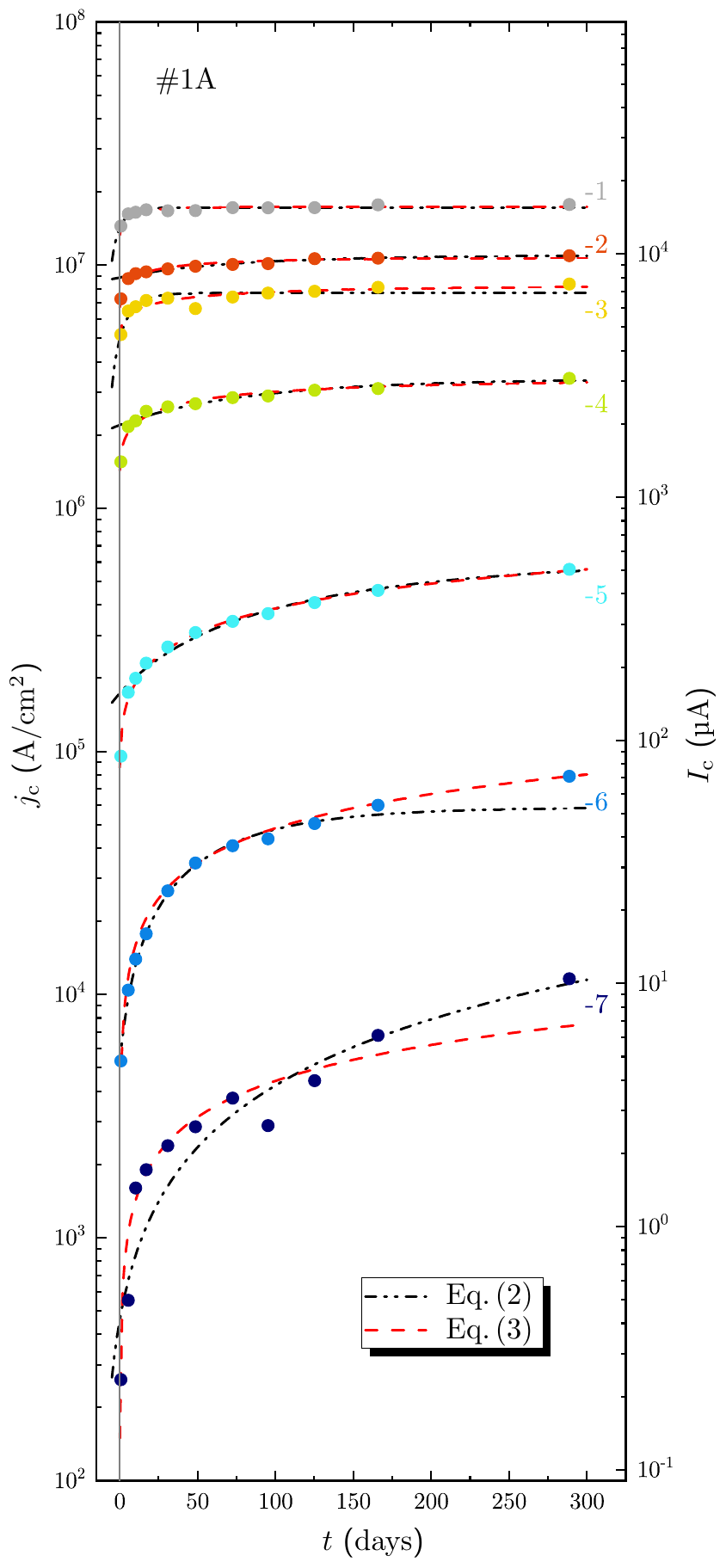}%{time-jc}
\end{center}
\caption{
Time dependence of the critical current density $\ls{j}{c}(t)$ (right axis: critical current $\ls{I}{c}$) for junctions 1 to 7 from \#1A written with different doses (see Tab.~\ref{tab:dose-values}) and stored at room temperature.
The dash-dotted and dashed lines show fits to the experimental data (symbols), obtained from Eq.~\eqref{eq:ExpDecay} and \eqref{eq:diffusionfit}, respectively.
}
\label{fig:time-jc}
\end{figure}

Here, we present results
%on the temporal evolution of the electric transport properties of
on He-FIB JJs which were stored at room temperature.
Figure \ref{fig:time-jc} shows the time dependence of the critical current density $\ls{j}{c}(t)$ of the seven JJs \#1A-1 to \#1A-7 located in structure \#1A over 160 days after irradiation.
Obviously, $\ls{j}{c}$ is largest for the JJ irradiated with the lowest dose (top curve in Fig.~\ref{fig:time-jc}) and smallest for the JJ irradiated with the highest dose (bottom curve in Fig.~\ref{fig:time-jc}).
We find a monotonous increase in $\ls{j}{c}(t)$, which tends to saturate for long enough times.
However, we also see a clear dose dependence in the evolution of $\ls{j}{c}(t)$.
For the lowest dose (junction \#1A-1), already after a few days $\ls{j}{c}$ saturates.
With increasing dose (junctions \#1A-2 to \#1A-4), the time until saturation of $\ls{j}{c}$ is reached steadily increases, and for the highest doses (junctions \#1A-5 to \#1A-7) $\ls{j}{c}(t)$ has not saturated at $t=288\units{days}$.

%%%%%%%%%%% Fig.4 %%%%%%%%%%%%%%%%%%%%%%%%%%%%%%%%%%%%%
%%%%%%%%%%% Fig.4 %%%%%%%%%%%%%%%%%%%%%%%%%%%%%%%%%%%%%

To fit the experimental data we used two different equations
\begin{align}
  \f{\ls{j}{c}}{t} &= \ls{j}{c,\infty} \cdot \left[ 1 - b \cdot \f{\exp}{-\frac{t}{\tau}} \right];\text{ and }
  \label{eq:ExpDecay}\\
  \f{\ls{j}{c}}{t} &= \ls{j}{c,\infty} \cdot \left[ 1 - b \cdot \f{\exp}{-\sqrt{\frac{t}{\tau}}} \right]
  \label{eq:diffusionfit}
\end{align}
Here $\ls{j}{c,\infty}$ is the saturation (equilibrium) value of the critical current, which is achieved for $t\to\infty$.
This parameter can be associated with unrecoverable damage in the barrier region induced by He-FIB irradiation.
The dimensionless parameter $b=(\ls{j}{c,\infty}-\ls{j}{c,0})/\ls{j}{c,\infty}$ (with $\ls{j}{c,0}\equiv\f{\ls{j}{c}}{t=0}$) describes the degree of non-equilibrium damage that was created by He-FIB irradiation and which can still be recovered with time (by annealing).
$b=0$ means that we are at equilibrium already at $t=0$ and there is nothing to recover.
$b=1$ means that the He-FIB irradiation initially completely suppresses the critical current density, i.e. $\ls{j}{c,0}=0$, but with time finite $\ls{j}{c}$ will appear.
The parameter $\tau$ describes the characteristic time scale (relaxation time) that governs relaxation of the system to the equilibrium state.
The two formulas are similar, but Eq.~\eqref{eq:ExpDecay} describes the changes in $\ls{j}{c}$ due to a \emph{constant flux} of oxygen atoms diffusing towards the barrier, while eq.~\eqref{eq:diffusionfit} describes the \emph{diffusion} of a limited amount of oxygen atoms towards the barrier.
Eq.~\eqref{eq:ExpDecay} assumes that the sample is placed in a oxygen atmosphere and the oxygen atoms diffuse into the barrier from the outside; instead eq.~\ref{eq:diffusionfit} assumes that oxygen atoms that have been displaced during He-FIB irradiation diffuse back similar to Einstein diffusion.
By using eq.~\eqref{eq:ExpDecay} and eq.~\eqref{eq:diffusionfit} to fit the data shown in Fig.~\ref{fig:time-jc}, we conclude that eq.~\eqref{eq:diffusionfit} allows to fit the experimental data more consistently at smaller times for devices irradiated with higher doses.

%%%%%%%%%%% Fig.5 %%%%%%%%%%%%%%%%%%%%%%%%%%%%%%%%%%%%%
\begin{figure}[!htb]
\begin{center}
\includegraphics*[width=1\columnwidth]{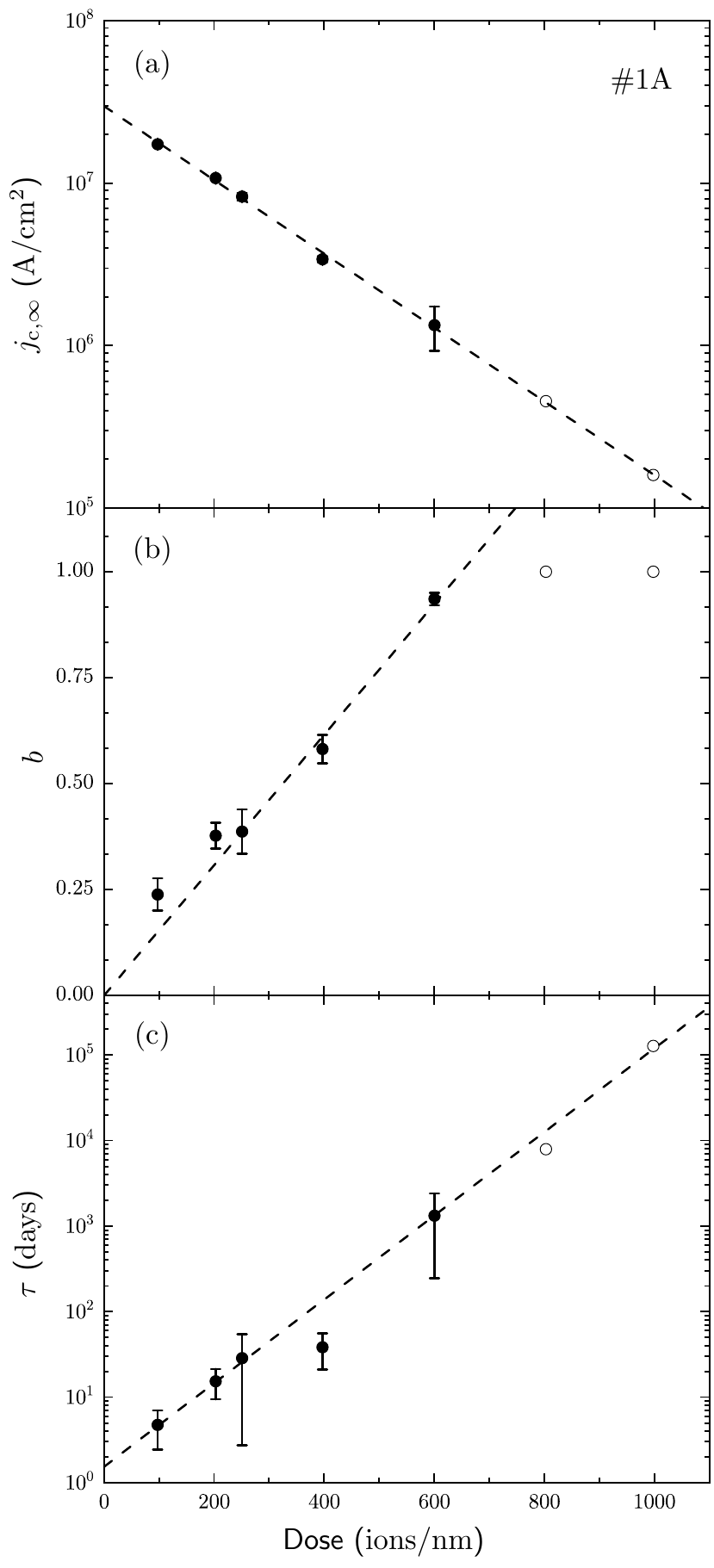}%{FitParameters}
\end{center}
\caption{
Dose dependence of the parameters $\ls{j}{c,\infty}$ (a), $b$ (b) and $\tau$ (c) extracted from the fits with eq.~\eqref{eq:diffusionfit} to $\ls{j}{c}(t)$ data shown in fig.~\ref{fig:time-jc}.
For the two highest dose values, the $\ls{j}{c,\infty}$ values (open symbols in (a)) have been fixed for the fitting procedure to obtain $b$ and $\tau$ \cite{note1}.
The confidence intervals provided by the fitting algorithm are shown as error bars.
}
  \label{fig:fitparams}
\end{figure}
%%%%%%%%%%% Fig.5 %%%%%%%%%%%%%%%%%%%%%%%%%%%%%%%%%%%%%

\begin{figure}[!htb]
\begin{center}
\includegraphics*[width=1\columnwidth]{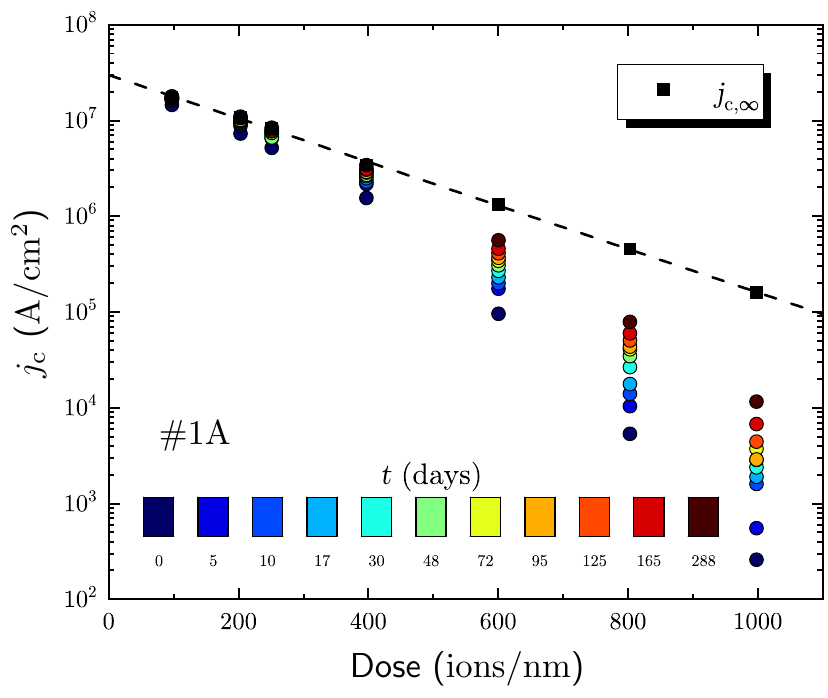}%{jc(D)}
\end{center}
\caption{
Experimentally determined critical current density $\ls{j}{c}$ vs.~irradiation dose $D$ for junctions from \#1A, measured at different times $t$ after irradiation (indicated by the color scale).
For comparison the fit parameter $\ls{j}{c,\infty}(D)$ is shown as black squares.
}
\label{fig:jc(D)}
\end{figure}

As a result of fitting with eq.~\eqref{eq:diffusionfit} we obtain the dose dependence of the fitting parameters $\ls{j}{c,\infty}$, $b$ and $\tau$ as shown in fig.~\ref{fig:fitparams}.
For $\ls{j}{c,\infty}$ we find an exponential decay with increasing dose (see fig.~\ref{fig:fitparams}(a)), i.e.,
\begin{equation}
  \f{\ls{j}{c,\infty}}{D} \approx \f{\ls{j}{c,\infty}}{0}\f{\exp}{-D/\ls{D}{0}}
  \label{Eq:jc.inf(D)}
\end{equation}
with $\f{\ls{j}{c,\infty}}{0}=3.2\times 10^7{\rm A/cm}^2$ and $\ls{D}{0}=191\units{ions/nm}$.
The value for $\f{\ls{j}{c,\infty}}{0}$ corresponds to the critical current density of unirradiated YBCO films at $4.2\units{K}$; this coincides quite well with the corresponding value given in Ref.~[\onlinecite{Mueller19}].
The value for the characteristic dose $\ls{D}{0}$ obtained in the present study is slightly higher than the value $\ls{D}{0}=130\units{ions/nm}$ found for the exponential scaling of $\f{\ls{j}{c}}{D}$ (measured at some finite $t$) for YBCO He-FIB JJs on STO or MgO substrates and significantly higher than $\ls{D}{0}=38\units{ions/nm}$ found for YBCO He-FIB JJs on a LSAT substrate \cite{Mueller19}.
The reason for the low value of $\ls{D}{0}=38\units{ions/nm}$ on LSAT in Ref.~[\onlinecite{Mueller19}] has not yet been clarified.
We note that we observe the tendency that for our YBCO He-FIB JJs produced on the same chip (typically on LSAT substrates), we always find an exponential decay of $\f{\ls{j}{c}}{D}$ with a well-defined $\ls{D}{0}$.
However, the value of $\ls{D}{0}$ can vary significantly from chip to chip.
Currently, it is not clear whether $\ls{D}{0}$ is related to and affected by the YBCO film quality or by the conditions in the helium ion microscope.
What is obvious from the study presented here, is the fact, that $\ls{D}{0}$ is also affected by the time $t$ (samples stored at room temparature) after He-FIB irradiation.
This is illustrated in fig.~\ref{fig:jc(D)}, which shows the same experimental data (dots) as in fig.~\ref{fig:time-jc}, but now plotted as  $\f{\ls{j}{c}}{D}$ for different values of $t$, together with  $\f{\ls{j}{c,\infty}}{D}$ (squares) from fig.~\ref{fig:fitparams}(a).
Clearly, with increasing $t$ the slope $|\partial\log\ls{j}{c}/\partial D|$ decreases, i.e., $\ls{D}{0}$ increases, approaching asymptotically the scaling of $\f{\ls{j}{c,\infty}}{D}$.
Thus, experimentally measured $\f{\ls{j}{c}}{D}$ tends to decrease the slope with time and becomes more straight.

%%%%%%%%%%% Fig.6 %%%%%%%%%%%%%%%%%%%%%%%%%%%%%%%%%%%%%
%%%%%%%%%%% Fig.6 %%%%%%%%%%%%%%%%%%%%%%%%%%%%%%%%%%%%%

The scaling of $b(D)$, shown in fig.~\ref{fig:fitparams}(b), is approximately linear for $D\lesssim 600\units{ions/nm}$ and saturates to $b\approx 1$ for larger $D$.
In the latter case, the permanent and non-equilibrium damage make $\ls{j}{c,0}=0$.
However, if one removes non-equilibrium damage by storing at room temperature, $\ls{j}{c}$ recovers to some finite value which is caused by permanent damage only.

Obviously, the characteristic time for recovery increases with $D$.
This is illustrated in  fig.~\ref{fig:fitparams}(c), which shows that the time constant $\tau$ increases exponentially with dose and can be described by
%
%%%%%%%%%%%%%%%%%%%%%%%%%%%%%%%
\begin{equation}
  \f{\tau}{D} \approx \ls{\tau}{0}\f{\exp}{D/\ls{D}{0,\tau}},
  \label{Eq:tau(D)}
\end{equation}
%%%%%%%%%%%%%%%%%%%%%%%%%%%%%%%
%
with $\ls{\tau}{0} = 1\units{day}$ and $\ls{D}{0,\tau} = 89\units{ions/nm}$.
Interestingly, the value of $\ls{D}{0,\tau}$ is not far from half of the value of $\ls{D}{0}$.
However, it is not clear whether or not this is just a coincidence.
Still, it is important to note that for typical doses required to create RCSJ-like Josephson junctions without an excess current, i.e., $D\gtrsim 500\units{ions/nm}$, the value of $\tau\gtrsim 100\units{days}$.
This means that the properties of junctions produced with such doses change over significant time scales of up to a year, if they are stored at room temperature.

%%%%%%%%%%% Fig.7 %%%%%%%%%%%%%%%%%%%%%%%%%%%%%%%%%%%%%
\begin{figure}[!htb]
\begin{center}
\includegraphics*[width=1\columnwidth]{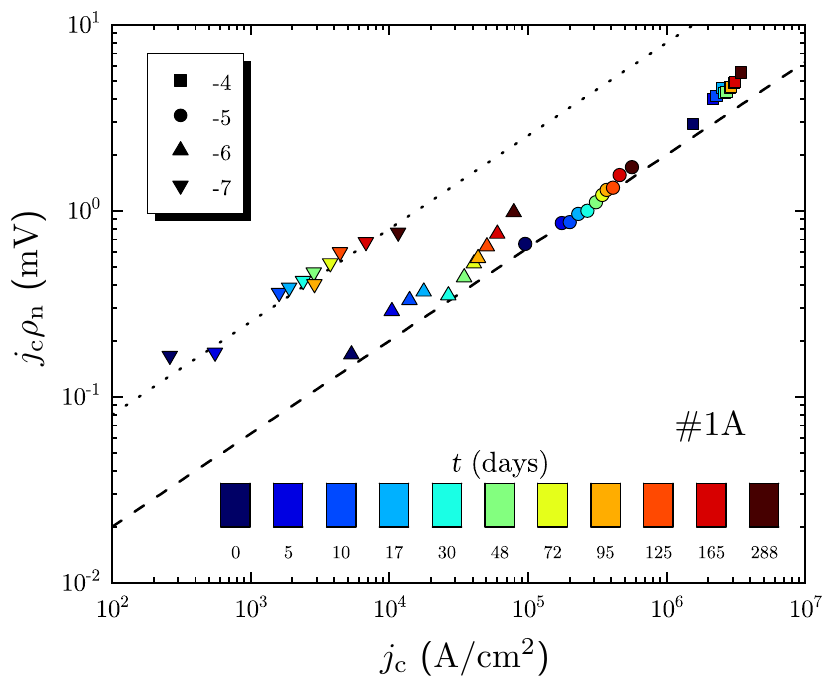}%{jc-jcrho}
\end{center}
\caption{
$\ls{j}{c}\ls{\rho}{n}$ vs $\ls{j}{c}$ for He-FIB JJs (from \#1A) stored at room temperature and measured at different times $t$ after irradiation (indicated by the color scale).
The dashed and dotted lines represent $\ls{j}{c}\ls{\rho}{n} = \ls{V}{c,1}\sqrt{\ls{j}{c}/\ls{j}{c,1}}$, with $\ls{V}{c,1}=2\units{mV}$ and $8\units{mV}$, respectively ($\ls{j}{c,1}=10^6\units{A/cm^2}$).
}
\label{fig:jc-jcrho}
\end{figure}
%%%%%%%%%%% Fig.7 %%%%%%%%%%%%%%%%%%%%%%%%%%%%%%%%%%%%%

In addition to $\f{\ls{j}{c}}{t}$ shown in fig.~\ref{fig:time-jc}, we also measured the time evolution $\f{\ls{R}{n}}{t}$ and calculated $\f{\ls{\rho}{n}}{t}$ (not shown).
This allows us to calculate $\f{\ls{V}{c}}{t}=\f{\ls{j}{c}}{t} \cdot \f{\ls{\rho}{n}}{t}$ and its evolution during annealing, see fig.~\ref{fig:jc-jcrho}.
In Ref.~[\onlinecite{Mueller19}] we showed that the exponential scaling of characteristic He-FIB JJ properties ($\ls{j}{c}$, $\ls{\rho}{n}=\ls{R}{n}wd$ and the characteristic voltage $\ls{V}{c} \equiv \ls{I}{c}\ls{R}{n}\equiv \ls{j}{c}\ls{\rho}{n}$) with $D$ can be described by the same values of $\ls{D}{0}$, indicating a universal scaling of $\ls{V}{c}$ with either $\ls{j}{c}$ or $\ls{\rho}{n}$, and we found an approximate scaling $\ls{V}{c}\approx\ls{V}{c,1}(\ls{j}{c}/\ls{j}{c,1})^{1/2}$.
%$\ls{V}{c}\propto\ls{j}{c}^{1/2}$.
%
Now, for the He-FIB JJs described here, we find the same scaling $\ls{V}{c}\propto\ls{j}{c}^{1/2}$, as shown in fig.~\ref{fig:jc-jcrho}.
Note that this is not only true for the junctions measured directly after irradiation, but also for junctions stored at room temperature and remeasured over the whole time of the experiment (up to $t=288\units{days}$), i.e., for the same junction, the data points move with $t$ along the $\ls{V}{c}\propto\ls{j}{c}^{1/2}$ line.
The data points for the He-FIB JJs \#1A-4, \#1A-5 and \#1A-6 follow the same scaling with $\ls{V}{c,1}=2\units{mV}$ and $\ls{j}{c,1}=10^6\units{A/cm^2}$ as reported in Ref.~[\onlinecite{Mueller19}].
However the data points for junction \#1A-7 with $D=1000\units{ions/nm}$ show a significant offset i.e., they also show the scaling $\ls{V}{c}\propto\ls{j}{c}^{1/2}$, but now with a factor of four larger $\ls{V}{c,1}=8\units{mV}$ and $\ls{j}{c,1}=10^6\units{A/cm^2}$.

\subsection{Temporal evolution of electric transport properties of He-FIB JJs that were annealed after irradiation}

%%%%%%%%%%% Fig.8 %%%%%%%%%%%%%%%%%%%%%%%%%%%%%%%%%%%%%
\begin{figure}[!htb]
\begin{center}
\includegraphics*[width=1\columnwidth]{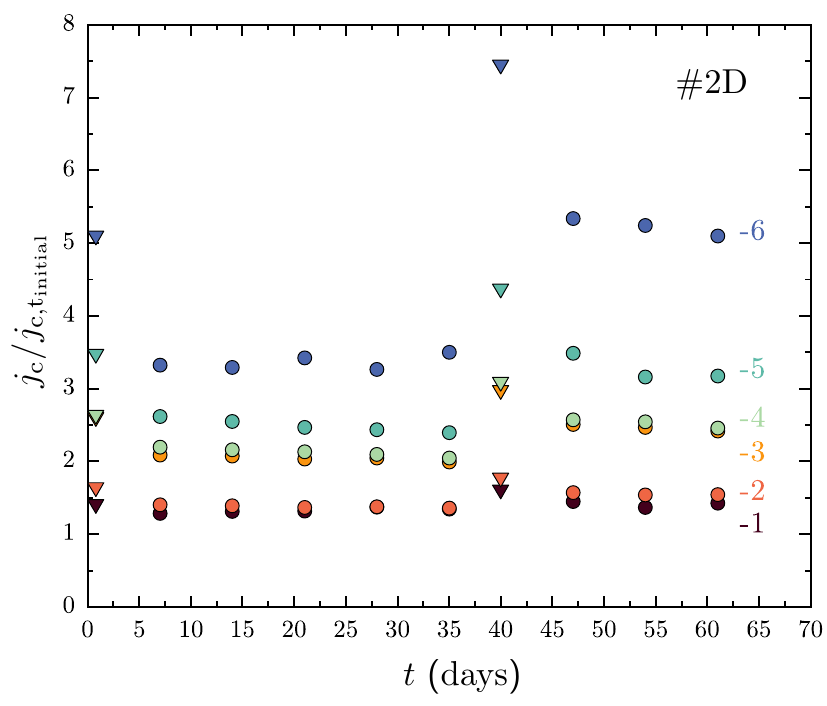}%{Anneal-normalized}
\end{center}
\caption{
Temporal evolution of the critical current density $\ls{j}{c}$ (normalized to $\ls{j}{c}$ at $t=\ls{t}{initial}$) of He-FIB JJs from \#2D, which were annealed at $90^\circ$C for $30\units{min}$.
\label{fig:anneal-jcnorm}
}
\end{figure}
%%%%%%%%%%% Fig.8 %%%%%%%%%%%%%%%%%%%%%%%%%%%%%%%%%%%%%

The temporal evolution of the critical current density of He-FIB JJs from structure \#2D is presented in fig.~\ref{fig:anneal-jcnorm}.
Those junctions have been annealed twice in oxygen ($\ls{p}{O_2}=950\units{mbar}$) for $30\units{min}$ at day 1 and day 41 ($j_c$ measurements on day 1 and 41 have been performed after annealing).
Annealing the junctions at day 1 caused an increase in $\ls{j}{c}$ and a decrease in $\ls{\rho}{n}$, respectively, just after the annealing step.
After one week the critical currents were reduced, but still maintained higher values than obtained directly after irradiation.
In the following weeks $\ls{j}{c}$ did not change significantly, indicating that the junctions have reached a quasi-stable state.
On day 41 we repeated the same annealing procedure with the already annealed junctions.
The temporal evolution of $\ls{j}{c}$ showed qualitatively the same behavior as after the first annealing step, however $\ls{j}{c}$ values are higher and there seems to appear now a short but finite relaxation time until another quasi-stable state has been reached.

%%%%%%%%%%% Fig.9 %%%%%%%%%%%%%%%%%%%%%%%%%%%%%%%%%%%%%
\begin{figure}[!htb]
\begin{center}
\includegraphics*[width=1\columnwidth]{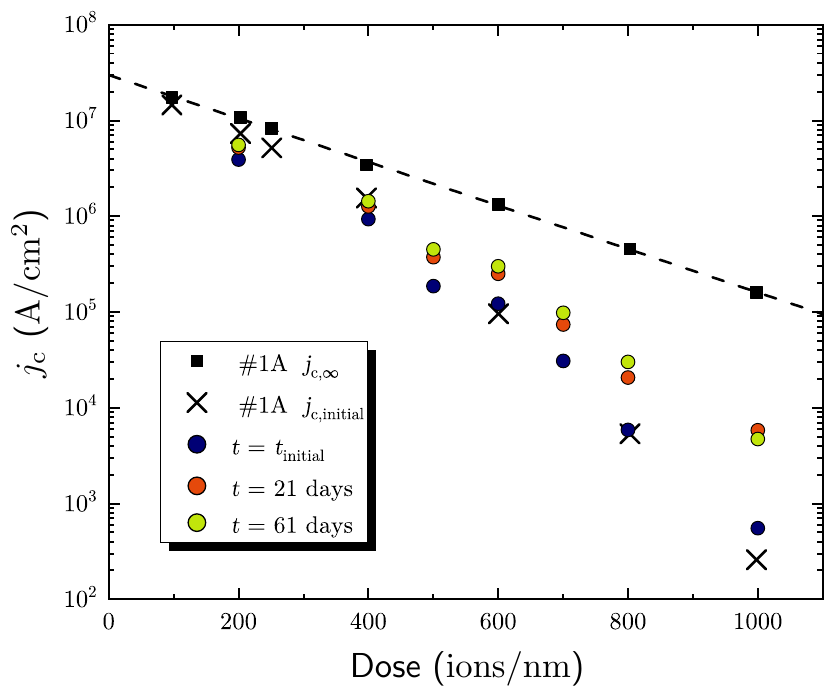}%{Anneal-jc(D)}
\end{center}
\caption{
Critical current density $\ls{j}{c}$ vs.~irradiation dose $D$ for He-FIB JJs from \#2D, measured at substantial times ($t=\ls{t}{initial}$, $21\units{days}$ and $61\units{days}$).
For comparison, we show results from JJs from \#1A -- stored at room temperature -- for the initial critical current density (black crosses)  and $\f{\ls{j}{c,\infty}}{D}$ (dashed line and  black sqares).
\label{fig:anneal-jcD}
}
\end{figure}
%%%%%%%%%%% Fig.9 %%%%%%%%%%%%%%%%%%%%%%%%%%%%%%%%%%%%%

In fig.~\ref{fig:anneal-jcD}, the data from fig.~\ref{fig:anneal-jcnorm} are plotted against the dose $D$ for substantial points in time,
namely after irradiation ($t=\ls{t}{initial}$) and for both quasi-stable states after relaxation following the first and second annealing step ($t=21$ and $61\units{days}$, respectively).
We find, that with comparable initial critical current densities (right after He-FIB irradiation), the first annealing step increased $\ls{j}{c}$ equivalently to roughly $100\units{days}$ of storing at room temperature, while the second annealing step produced another increase in $\ls{j}{c}$, albeit at a moderate level.
Most importantly, the junctions after annealing do not exhibit a significant further change in $\ls{j}{c}$ with time (c.f.~Fig.~\ref{fig:anneal-jcnorm}), which one would expect if following the time evolution shown in fig.~\ref{fig:time-jc}, but remained unchanged in their electric transport properties.
This clearly indicates that moderate thermal annealing after He-FIB irradiation clearly stabilizes the irradiation-induced defect structure.

\begin{figure}[!htb]
\begin{center}
\includegraphics*[width=1\columnwidth]{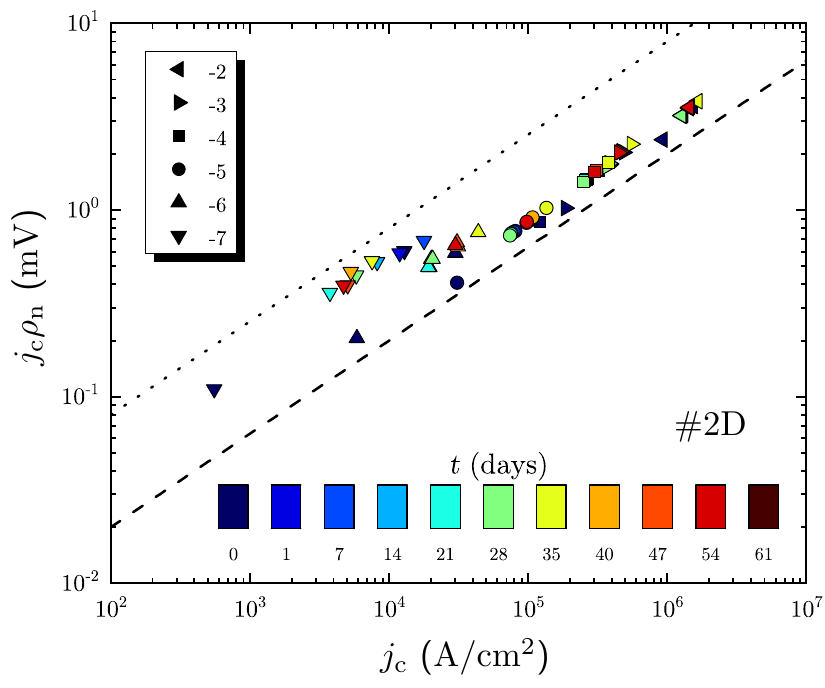}%{jc-jcrho-anneal}
\end{center}
\caption{
$\ls{j}{c}\ls{\rho}{n}$ vs $\ls{j}{c}$ for annealed He-FIB JJs (from \#2D) under high oxygen pressure.
The dashed and dotted lines represent $\ls{j}{c}\ls{\rho}{n} =  \ls{V}{c,1}\sqrt{\ls{j}{c}/\ls{j}{c,1}}$, with $\ls{V}{c,1}=2\units{mV}$ and $8\units{mV}$, respectively ($\ls{j}{c,1}=10^6\units{A/cm^2}$).
}
  \label{fig:jc-jcrho-anneal}
\end{figure}

Similar to the graph in fig.~\ref{fig:jc-jcrho} the overall scaling $V_c(j_c)$ for the annealed junctions is shown in fig.~\ref{fig:jc-jcrho-anneal}.
As the junctions stored at room temperature, the annealed junctions follow the same $\ls{V}{c}\propto\ls{j}{c}^{1/2}$ scaling indicated by the dotted and dashed lines.
In addition one can see that the higher dose junctions are shifted towards higher characteristic voltages for the same critical current density.

For comparison, and to rule out the possibility that the increase in $\ls{j}{c}$ shortly after annealing is caused by the high oxygen pressure in the annealing chamber during the treatment, we performed annealing of He-FIB JJs from structure \#2E at $\ls{p}{O_2}< 0.1\units{mbar}$.
Here, we could not observe a substantial difference of the temporal evolution of $\ls{j}{c}$ as compared to junctions annealed at $\ls{p}{O_2}=950\units{mbar}$.
This observation indicates, that the restoring of critical current density with annealing (or also simply by storing at room temperature) is due to the diffusion of oxygen atoms that have been displaced (probably to interstitial sites) by He-FIB irradiation and that are diffusing back to lattice sites.
Obviously, the annealing process significantly speeds up the relaxation (back diffusion) to a quasi-stable state with stable electric transport properties.

%%%%%%%%%%% Fig.10 %%%%%%%%%%%%%%%%%%%%%%%%%%%%%%%%%%%%%
%%%%%%%%%%% Fig.10 %%%%%%%%%%%%%%%%%%%%%%%%%%%%%%%%%%%%%

%%%%%%%%%%%%%%%%%%%%%%%%%%%%%%%%%
\section{Conclusions}
\label{sec:conclusions}
%%%%%%%%%%%%%%%%%%%%%%%%%%%%%%%%%

We fabricated Josephson barriers in prepatterned $3\units{\mu m}$ wide microbridges of epitaxially grown YBCO thin films by scanning a focused He ion beam across the bridges, which locally modifies the superconducting properties of the thin film.
In detail, we studied the temporal evolution of characteristic junction properties for devices stored at room temperature and found that this process can be best described by a diffusion based model, corresponding to a limited amount of oxygen diffusing back towards the barrier.
Moreover, fitting of the experimentally observed temporal evolution of critical current density with time, allowed us to extract the dose dependence of the fitting parameters $\ls{j}{c,\infty},b$ and $\tau$.
For high quality junctions, corresponding to doses in the range of $600-800\units{ions/nm}$, the relaxation time is typically $\tau\sim 100-1000\units{days}$, exceeding any feasible timescale to wait for the equilibrium state.
The analysis of the characteristic voltages $\ls{V}{c}$ of the fabricated devices yield a scaling $\ls{V}{c}\propto\sqrt{\ls{j}{c}}$, also reported in previous work~\cite{Mueller19}, but now this scaling holds true even for junctions stored at room temperature over the whole time of the experiment.

Furthermore, we examined the behavior of devices when treated with an annealing step at high oxygen pressures.
Directly after the treatment the critical current density $\ls{j}{c}$ increased before relaxing into a quasi-stable state within the first week.
A repeated annealing step increased $\ls{j}{c}$ only marginally, but indicated a slightly larger relaxation time of roughly two weeks.
In comparison to junctions stored at room temperature, the annealing process accelerated the increase in $\ls{j}{c}$ to roughly an equivalent of $100\units{days}$ storing at room temperature.
Additionally, for annealing at low oxygen pressures we did not observe any substantial differences in the behavior of the annealed devices annealed at high oxygen pressure.
This indicates that predominantly the oxygen atoms displaced during He-FIB irradiation are diffusing back towards the original lattice sites.

Although some enigma remain to be examined in further studies, we revealed how characteristic properties of He-FIB JJs in YBCO evolve with time when stored at room temperature and confirmed a strategy to stabilize their properties.

%%%%%%%%%%%%%%%%%%%%%%%%%%%%%%%%%
\section*{Acknowledgments}
%%%%%%%%%%%%%%%%%%%%%%%%%%%%%%%%%

We gratefully acknowledge technical support by M.~Turad, R.~L\"offler (LISA$^+$), C.~Back and R.~Schilling.
This work was supported by the Deutsche Forschungsgemeinschaft (DFG) (GO 1106/6-1), by the European Commission under H2020 FET Open grant ‘FIBsuperProbes’ (Grant No. 892427) and by the COST actions FIT4NANO (CA19140) and SUPERQUMAP (CA21144).

\bibliography{HIM-JJs}

%merlin.mbs apsrev4-1.bst 2010-07-25 4.21a (PWD, AO, DPC) hacked
%Control: key (0)
%Control: author (0) dotless jnrlst
%Control: editor formatted (1) identically to author
%Control: production of article title (0) allowed
%Control: page (1) range
%Control: year (0) verbatim
%Control: production of eprint (0) enabled
\begin{thebibliography}{26}%
\makeatletter
\providecommand \@ifxundefined [1]{%
 \@ifx{#1\undefined}
}%
\providecommand \@ifnum [1]{%
 \ifnum #1\expandafter \@firstoftwo
 \else \expandafter \@secondoftwo
 \fi
}%
\providecommand \@ifx [1]{%
 \ifx #1\expandafter \@firstoftwo
 \else \expandafter \@secondoftwo
 \fi
}%
\providecommand \natexlab [1]{#1}%
\providecommand \enquote  [1]{``#1''}%
\providecommand \bibnamefont  [1]{#1}%
\providecommand \bibfnamefont [1]{#1}%
\providecommand \citenamefont [1]{#1}%
\providecommand \href@noop [0]{\@secondoftwo}%
\providecommand \href [0]{\begingroup \@sanitize@url \@href}%
\providecommand \@href[1]{\@@startlink{#1}\@@href}%
\providecommand \@@href[1]{\endgroup#1\@@endlink}%
\providecommand \@sanitize@url [0]{\catcode `\\12\catcode `\$12\catcode
  `\&12\catcode `\#12\catcode `\^12\catcode `\_12\catcode `\%12\relax}%
\providecommand \@@startlink[1]{}%
\providecommand \@@endlink[0]{}%
\providecommand \url  [0]{\begingroup\@sanitize@url \@url }%
\providecommand \@url [1]{\endgroup\@href {#1}{\urlprefix }}%
\providecommand \urlprefix  [0]{URL }%
\providecommand \Eprint [0]{\href }%
\providecommand \doibase [0]{http://dx.doi.org/}%
\providecommand \selectlanguage [0]{\@gobble}%
\providecommand \bibinfo  [0]{\@secondoftwo}%
\providecommand \bibfield  [0]{\@secondoftwo}%
\providecommand \translation [1]{[#1]}%
\providecommand \BibitemOpen [0]{}%
\providecommand \bibitemStop [0]{}%
\providecommand \bibitemNoStop [0]{.\EOS\space}%
\providecommand \EOS [0]{\spacefactor3000\relax}%
\providecommand \BibitemShut  [1]{\csname bibitem#1\endcsname}%
\let\auto@bib@innerbib\@empty
%</preamble>
\bibitem [{\citenamefont {Koelle}\ \emph {et~al.}(1999)\citenamefont {Koelle},
  \citenamefont {Kleiner}, \citenamefont {Ludwig}, \citenamefont {Dantsker},\
  and\ \citenamefont {Clarke}}]{Koelle99}%
  \BibitemOpen
  \bibfield  {author} {\bibinfo {author} {\bibfnamefont {D.}~\bibnamefont
  {Koelle}}, \bibinfo {author} {\bibfnamefont {R.}~\bibnamefont {Kleiner}},
  \bibinfo {author} {\bibfnamefont {F.}~\bibnamefont {Ludwig}}, \bibinfo
  {author} {\bibfnamefont {E.}~\bibnamefont {Dantsker}}, \ and\ \bibinfo
  {author} {\bibfnamefont {John}\ \bibnamefont {Clarke}},\ }\bibfield  {title}
  {\enquote {\bibinfo {title} {High-transition-temperature superconducting
  quantum interference devices},}\ }\href {\doibase 10.1103/RevModPhys.71.631}
  {\bibfield  {journal} {\bibinfo  {journal} {Rev. Mod. Phys.}\ }\textbf
  {\bibinfo {volume} {71}},\ \bibinfo {pages} {631--686} (\bibinfo {year}
  {1999})}\BibitemShut {NoStop}%
\bibitem [{\citenamefont {Ludwig}\ \emph {et~al.}(1995)\citenamefont {Ludwig},
  \citenamefont {Koelle}, \citenamefont {Dantsker}, \citenamefont {Nemeth},
  \citenamefont {Miklich}, \citenamefont {Clarke},\ and\ \citenamefont
  {Thomson}}]{Ludwig95}%
  \BibitemOpen
  \bibfield  {author} {\bibinfo {author} {\bibfnamefont {F.}~\bibnamefont
  {Ludwig}}, \bibinfo {author} {\bibfnamefont {D.}~\bibnamefont {Koelle}},
  \bibinfo {author} {\bibfnamefont {E.}~\bibnamefont {Dantsker}}, \bibinfo
  {author} {\bibfnamefont {D.T.}\ \bibnamefont {Nemeth}}, \bibinfo {author}
  {\bibfnamefont {A.H.}\ \bibnamefont {Miklich}}, \bibinfo {author}
  {\bibfnamefont {J.}~\bibnamefont {Clarke}}, \ and\ \bibinfo {author}
  {\bibfnamefont {R.E.}\ \bibnamefont {Thomson}},\ }\bibfield  {title}
  {\enquote {\bibinfo {title} {Low noise
  {YBa$_2$Cu$_3$O$_{7-x}$-SrTiO$_3$-YBa$_2$Cu$_3$O$_{7-x}$} multilayers for
  improved magnetometers},}\ }\href {\doibase 10.1063/1.114217} {\bibfield
  {journal} {\bibinfo  {journal} {Appl. Phys. Lett.}\ }\textbf {\bibinfo
  {volume} {66}},\ \bibinfo {pages} {373--375} (\bibinfo {year}
  {1995})}\BibitemShut {NoStop}%
\bibitem [{\citenamefont {Hilgenkamp}\ and\ \citenamefont
  {Mannhart}(2002)}]{Hilgenkamp02}%
  \BibitemOpen
  \bibfield  {author} {\bibinfo {author} {\bibfnamefont {H.}~\bibnamefont
  {Hilgenkamp}}\ and\ \bibinfo {author} {\bibfnamefont {J.}~\bibnamefont
  {Mannhart}},\ }\bibfield  {title} {\enquote {\bibinfo {title} {Grain
  boundaries in high-{$T_c$} superconductors},}\ }\href {\doibase
  10.1103/RevModPhys.74.485} {\bibfield  {journal} {\bibinfo  {journal} {Rev.
  Mod. Phys.}\ }\textbf {\bibinfo {volume} {74}},\ \bibinfo {pages} {485}
  (\bibinfo {year} {2002})}\BibitemShut {NoStop}%
\bibitem [{\citenamefont {Gross}\ \emph {et~al.}(1997)\citenamefont {Gross},
  \citenamefont {Alff}, \citenamefont {Beck}, \citenamefont {Froehlich},
  \citenamefont {Koelle},\ and\ \citenamefont {Marx}}]{Gross97}%
  \BibitemOpen
  \bibfield  {author} {\bibinfo {author} {\bibfnamefont {R.}~\bibnamefont
  {Gross}}, \bibinfo {author} {\bibfnamefont {L.}~\bibnamefont {Alff}},
  \bibinfo {author} {\bibfnamefont {A.}~\bibnamefont {Beck}}, \bibinfo {author}
  {\bibfnamefont {O.M.}\ \bibnamefont {Froehlich}}, \bibinfo {author}
  {\bibfnamefont {D.}~\bibnamefont {Koelle}}, \ and\ \bibinfo {author}
  {\bibfnamefont {A.}~\bibnamefont {Marx}},\ }\bibfield  {title} {\enquote
  {\bibinfo {title} {Physics and technology of high temperature superconducting
  {J}osephson junctions},}\ }\href {\doibase 10.1109/77.621919} {\bibfield
  {journal} {\bibinfo  {journal} {IEEE Trans. Appl. Supercond.}\ }\textbf
  {\bibinfo {volume} {7}},\ \bibinfo {pages} {2929--2935} (\bibinfo {year}
  {1997})}\BibitemShut {NoStop}%
\bibitem [{\citenamefont {Bergeal}\ \emph {et~al.}(2007)\citenamefont
  {Bergeal}, \citenamefont {Lesueur}, \citenamefont {Sirena}, \citenamefont
  {Faini}, \citenamefont {Aprili}, \citenamefont {Contour},\ and\ \citenamefont
  {Leridon}}]{Bergeal2007}%
  \BibitemOpen
  \bibfield  {author} {\bibinfo {author} {\bibfnamefont {N.}~\bibnamefont
  {Bergeal}}, \bibinfo {author} {\bibfnamefont {J.}~\bibnamefont {Lesueur}},
  \bibinfo {author} {\bibfnamefont {M.}~\bibnamefont {Sirena}}, \bibinfo
  {author} {\bibfnamefont {G.}~\bibnamefont {Faini}}, \bibinfo {author}
  {\bibfnamefont {M.}~\bibnamefont {Aprili}}, \bibinfo {author} {\bibfnamefont
  {J.~P.}\ \bibnamefont {Contour}}, \ and\ \bibinfo {author} {\bibfnamefont
  {B.}~\bibnamefont {Leridon}},\ }\bibfield  {title} {\enquote {\bibinfo
  {title} {Using ion irradiation to make high-{$T_c$} {J}osephson junctions},}\
  }\href {\doibase 10.1063/1.2796105} {\bibfield  {journal} {\bibinfo
  {journal} {J.~Appl.~Phys.}\ }\textbf {\bibinfo {volume} {102}},\ \bibinfo
  {pages} {083903} (\bibinfo {year} {2007})}\BibitemShut {NoStop}%
\bibitem [{\citenamefont {Kang}\ \emph {et~al.}(2002)\citenamefont {Kang},
  \citenamefont {Burnell}, \citenamefont {Lloyd}, \citenamefont {Speaks},
  \citenamefont {Peng}, \citenamefont {Jeynes}, \citenamefont {Webb},
  \citenamefont {Yun}, \citenamefont {Moon}, \citenamefont {Oh}, \citenamefont
  {Tarte}, \citenamefont {Moore},\ and\ \citenamefont {Blamire}}]{Kang2002}%
  \BibitemOpen
  \bibfield  {author} {\bibinfo {author} {\bibfnamefont {D.-J.}\ \bibnamefont
  {Kang}}, \bibinfo {author} {\bibfnamefont {G.}~\bibnamefont {Burnell}},
  \bibinfo {author} {\bibfnamefont {S.~J.}\ \bibnamefont {Lloyd}}, \bibinfo
  {author} {\bibfnamefont {R.~S.}\ \bibnamefont {Speaks}}, \bibinfo {author}
  {\bibfnamefont {N.~H.}\ \bibnamefont {Peng}}, \bibinfo {author}
  {\bibfnamefont {C.}~\bibnamefont {Jeynes}}, \bibinfo {author} {\bibfnamefont
  {R.}~\bibnamefont {Webb}}, \bibinfo {author} {\bibfnamefont {J.~H.}\
  \bibnamefont {Yun}}, \bibinfo {author} {\bibfnamefont {S.~H.}\ \bibnamefont
  {Moon}}, \bibinfo {author} {\bibfnamefont {B.}~\bibnamefont {Oh}}, \bibinfo
  {author} {\bibfnamefont {E.~J.}\ \bibnamefont {Tarte}}, \bibinfo {author}
  {\bibfnamefont {D.~F.}\ \bibnamefont {Moore}}, \ and\ \bibinfo {author}
  {\bibfnamefont {M.~G.}\ \bibnamefont {Blamire}},\ }\bibfield  {title}
  {\enquote {\bibinfo {title} {Realization and properties of
  {YBa$_2$Cu$_3$O$_{7-\delta}$} {J}osephson junctions by metal masked ion
  damage technique},}\ }\href {\doibase 10.1063/1.1446998} {\bibfield
  {journal} {\bibinfo  {journal} {Appl.~Phys.~Lett.}\ }\textbf {\bibinfo
  {volume} {80}},\ \bibinfo {pages} {814--816} (\bibinfo {year}
  {2002})}\BibitemShut {NoStop}%
\bibitem [{\citenamefont {Tolpygo}\ \emph {et~al.}(1993)\citenamefont
  {Tolpygo}, \citenamefont {Shokhor}, \citenamefont {Nadgorny}, \citenamefont
  {Lin}, \citenamefont {Gurvitch}, \citenamefont {Bourdillon}, \citenamefont
  {Hou},\ and\ \citenamefont {Phillips}}]{Tolpygo93}%
  \BibitemOpen
  \bibfield  {author} {\bibinfo {author} {\bibfnamefont {S.~K.}\ \bibnamefont
  {Tolpygo}}, \bibinfo {author} {\bibfnamefont {S.}~\bibnamefont {Shokhor}},
  \bibinfo {author} {\bibfnamefont {B.}~\bibnamefont {Nadgorny}}, \bibinfo
  {author} {\bibfnamefont {J.-Y.}\ \bibnamefont {Lin}}, \bibinfo {author}
  {\bibfnamefont {M.}~\bibnamefont {Gurvitch}}, \bibinfo {author}
  {\bibfnamefont {A.}~\bibnamefont {Bourdillon}}, \bibinfo {author}
  {\bibfnamefont {S.~Y.}\ \bibnamefont {Hou}}, \ and\ \bibinfo {author}
  {\bibfnamefont {Julia~M.}\ \bibnamefont {Phillips}},\ }\bibfield  {title}
  {\enquote {\bibinfo {title} {High quality {YBa$_2$Cu$_3$O$_7$} {J}osephson
  junctions made by direct electron beam writing},}\ }\href {\doibase
  10.1063/1.110688} {\bibfield  {journal} {\bibinfo  {journal} {Appl. Phys.
  Lett.}\ }\textbf {\bibinfo {volume} {63}},\ \bibinfo {pages} {1696--1698}
  (\bibinfo {year} {1993})}\BibitemShut {NoStop}%
\bibitem [{\citenamefont {Booij}\ \emph {et~al.}(1997)\citenamefont {Booij},
  \citenamefont {Pauza}, \citenamefont {Tarte}, \citenamefont {Moore},\ and\
  \citenamefont {Blamire}}]{Booij1997}%
  \BibitemOpen
  \bibfield  {author} {\bibinfo {author} {\bibfnamefont {W.~E.}\ \bibnamefont
  {Booij}}, \bibinfo {author} {\bibfnamefont {A.~J.}\ \bibnamefont {Pauza}},
  \bibinfo {author} {\bibfnamefont {E.~J.}\ \bibnamefont {Tarte}}, \bibinfo
  {author} {\bibfnamefont {D.~F.}\ \bibnamefont {Moore}}, \ and\ \bibinfo
  {author} {\bibfnamefont {M.~G.}\ \bibnamefont {Blamire}},\ }\bibfield
  {title} {\enquote {\bibinfo {title} {Proximity coupling in high-$t_c$
  {J}osephson junctions produced by focused electron beam irradiation},}\
  }\href {\doibase 10.1103/physrevb.55.14600} {\bibfield  {journal} {\bibinfo
  {journal} {Phys.~Rev.~B}\ }\textbf {\bibinfo {volume} {55}},\ \bibinfo
  {pages} {14600--14609} (\bibinfo {year} {1997})}\BibitemShut {NoStop}%
\bibitem [{\citenamefont {Cybart}\ \emph {et~al.}(2015)\citenamefont {Cybart},
  \citenamefont {Cho}, \citenamefont {Wong}, \citenamefont {Wehlin},
  \citenamefont {Ma}, \citenamefont {Huynh},\ and\ \citenamefont
  {Dynes}}]{Cybart15}%
  \BibitemOpen
  \bibfield  {author} {\bibinfo {author} {\bibfnamefont {S.~A.}\ \bibnamefont
  {Cybart}}, \bibinfo {author} {\bibfnamefont {E.~Y.}\ \bibnamefont {Cho}},
  \bibinfo {author} {\bibfnamefont {J.~T.}\ \bibnamefont {Wong}}, \bibinfo
  {author} {\bibfnamefont {B.~H.}\ \bibnamefont {Wehlin}}, \bibinfo {author}
  {\bibfnamefont {M.~K.}\ \bibnamefont {Ma}}, \bibinfo {author} {\bibfnamefont
  {C.}~\bibnamefont {Huynh}}, \ and\ \bibinfo {author} {\bibfnamefont {R.~C.}\
  \bibnamefont {Dynes}},\ }\bibfield  {title} {\enquote {\bibinfo {title} {Nano
  {J}osephson superconducting tunnel junctions in {YBa$_2$Cu$_3$O$_{7-\delta}$}
  directly patterned with a focused helium ion beam},}\ }\href {\doibase
  10.1038/nnano.2015.76} {\bibfield  {journal} {\bibinfo  {journal} {Nat.
  Nano}\ }\textbf {\bibinfo {volume} {10}},\ \bibinfo {pages} {598--602}
  (\bibinfo {year} {2015})}\BibitemShut {NoStop}%
\bibitem [{\citenamefont {Cho}\ \emph {et~al.}(2018)\citenamefont {Cho},
  \citenamefont {Zhou}, \citenamefont {Cho},\ and\ \citenamefont
  {Cybart}}]{Cho18}%
  \BibitemOpen
  \bibfield  {author} {\bibinfo {author} {\bibfnamefont {E.~Y.}\ \bibnamefont
  {Cho}}, \bibinfo {author} {\bibfnamefont {Y.~W.}\ \bibnamefont {Zhou}},
  \bibinfo {author} {\bibfnamefont {J.~Y.}\ \bibnamefont {Cho}}, \ and\
  \bibinfo {author} {\bibfnamefont {S.~A.}\ \bibnamefont {Cybart}},\ }\bibfield
   {title} {\enquote {\bibinfo {title} {Superconducting nano {J}osephson
  junctions patterned with a focused helium ion beam},}\ }\href {\doibase
  10.1063/1.5042105} {\bibfield  {journal} {\bibinfo  {journal} {Appl. Phys.
  Lett.}\ }\textbf {\bibinfo {volume} {113}},\ \bibinfo {pages} {022604}
  (\bibinfo {year} {2018})}\BibitemShut {NoStop}%
\bibitem [{\citenamefont {M\"{u}ller}\ \emph {et~al.}(2019)\citenamefont
  {M\"{u}ller}, \citenamefont {Karrer}, \citenamefont {Limberger},
  \citenamefont {Becker}, \citenamefont {Schr\"{o}ppel}, \citenamefont
  {Burkhardt}, \citenamefont {Kleiner}, \citenamefont {Goldobin},\ and\
  \citenamefont {Koelle}}]{Mueller19}%
  \BibitemOpen
  \bibfield  {author} {\bibinfo {author} {\bibfnamefont {B.}~\bibnamefont
  {M\"{u}ller}}, \bibinfo {author} {\bibfnamefont {M.}~\bibnamefont {Karrer}},
  \bibinfo {author} {\bibfnamefont {F.}~\bibnamefont {Limberger}}, \bibinfo
  {author} {\bibfnamefont {M.}~\bibnamefont {Becker}}, \bibinfo {author}
  {\bibfnamefont {B.}~\bibnamefont {Schr\"{o}ppel}}, \bibinfo {author}
  {\bibfnamefont {C.~J.}\ \bibnamefont {Burkhardt}}, \bibinfo {author}
  {\bibfnamefont {R.}~\bibnamefont {Kleiner}}, \bibinfo {author} {\bibfnamefont
  {E.}~\bibnamefont {Goldobin}}, \ and\ \bibinfo {author} {\bibfnamefont
  {D.}~\bibnamefont {Koelle}},\ }\bibfield  {title} {\enquote {\bibinfo {title}
  {Josephson junctions and {SQUIDs} created by focused helium-ion-beam
  irradiation of {YBa$_2$Cu$_3$O$_7$}},}\ }\href {\doibase
  10.1103/PhysRevApplied.11.044082} {\bibfield  {journal} {\bibinfo  {journal}
  {Phys. Rev. Appl.}\ }\textbf {\bibinfo {volume} {11}},\ \bibinfo {pages}
  {044082} (\bibinfo {year} {2019})}\BibitemShut {NoStop}%
\bibitem [{\citenamefont {Cou\"{e}do}\ \emph {et~al.}(2020)\citenamefont
  {Cou\"{e}do}, \citenamefont {Amari}, \citenamefont {Feuillet-Palma},
  \citenamefont {Ulysse}, \citenamefont {Srivasta}, \citenamefont {Singh},
  \citenamefont {Bergeal},\ and\ \citenamefont {Leseur}}]{Couedo20}%
  \BibitemOpen
  \bibfield  {author} {\bibinfo {author} {\bibfnamefont {F.}~\bibnamefont
  {Cou\"{e}do}}, \bibinfo {author} {\bibfnamefont {P.}~\bibnamefont {Amari}},
  \bibinfo {author} {\bibfnamefont {C.}~\bibnamefont {Feuillet-Palma}},
  \bibinfo {author} {\bibfnamefont {C.}~\bibnamefont {Ulysse}}, \bibinfo
  {author} {\bibfnamefont {Y.~K.}\ \bibnamefont {Srivasta}}, \bibinfo {author}
  {\bibfnamefont {R.}~\bibnamefont {Singh}}, \bibinfo {author} {\bibfnamefont
  {N.}~\bibnamefont {Bergeal}}, \ and\ \bibinfo {author} {\bibfnamefont
  {J.}~\bibnamefont {Leseur}},\ }\bibfield  {title} {\enquote {\bibinfo {title}
  {Dynamic properties of high-{$T_c$} superconducting nano-junctions made with
  a focused helium ion beam},}\ }\href {\doibase 10.1088/2053-1583/acb4a8}
  {\bibfield  {journal} {\bibinfo  {journal} {Sci. Rep.}\ }\textbf {\bibinfo
  {volume} {10}},\ \bibinfo {pages} {021001} (\bibinfo {year}
  {2020})}\BibitemShut {NoStop}%
\bibitem [{\citenamefont {Chen}\ \emph {et~al.}(2022)\citenamefont {Chen},
  \citenamefont {Li}, \citenamefont {Zhu}, \citenamefont {Xu}, \citenamefont
  {T.~Xu}, \citenamefont {Cai}, \citenamefont {Wang}, \citenamefont {Lu},
  \citenamefont {Zhang},\ and\ \citenamefont {Ma}}]{Chen22}%
  \BibitemOpen
  \bibfield  {author} {\bibinfo {author} {\bibfnamefont {Z.}~\bibnamefont
  {Chen}}, \bibinfo {author} {\bibfnamefont {Y.}~\bibnamefont {Li}}, \bibinfo
  {author} {\bibfnamefont {R.}~\bibnamefont {Zhu}}, \bibinfo {author}
  {\bibfnamefont {J.}~\bibnamefont {Xu}}, \bibinfo {author} {\bibfnamefont
  {D.~Yin}\ \bibnamefont {T.~Xu}}, \bibinfo {author} {\bibfnamefont
  {X.}~\bibnamefont {Cai}}, \bibinfo {author} {\bibfnamefont {Y.}~\bibnamefont
  {Wang}}, \bibinfo {author} {\bibfnamefont {J.}~\bibnamefont {Lu}}, \bibinfo
  {author} {\bibfnamefont {Y.}~\bibnamefont {Zhang}}, \ and\ \bibinfo {author}
  {\bibfnamefont {P.}~\bibnamefont {Ma}},\ }\bibfield  {title} {\enquote
  {\bibinfo {title} {High-temperature superconducting
  {YBa$_2$Cu$_3$O$_{7-\delta}$} {J}osephson junction fabricated with a focused
  helium ion beam},}\ }\href {\doibase 10.1088/0256-307X/39/7/077402}
  {\bibfield  {journal} {\bibinfo  {journal} {Chin. Phys. Lett.}\ }\textbf
  {\bibinfo {volume} {39}},\ \bibinfo {pages} {077402} (\bibinfo {year}
  {2022})}\BibitemShut {NoStop}%
\bibitem [{\citenamefont {Merino}\ \emph {et~al.}(2023)\citenamefont {Merino},
  \citenamefont {Seifert}, \citenamefont {Retamal}, \citenamefont {Mech},
  \citenamefont {Taniguchi}, \citenamefont {Watanabe}, \citenamefont
  {Kadowaki}, \citenamefont {Hadfield},\ and\ \citenamefont
  {Efetov}}]{Merino23}%
  \BibitemOpen
  \bibfield  {author} {\bibinfo {author} {\bibfnamefont {R.~L.}\ \bibnamefont
  {Merino}}, \bibinfo {author} {\bibfnamefont {P.}~\bibnamefont {Seifert}},
  \bibinfo {author} {\bibfnamefont {J.~D.}\ \bibnamefont {Retamal}}, \bibinfo
  {author} {\bibfnamefont {R.~K.}\ \bibnamefont {Mech}}, \bibinfo {author}
  {\bibfnamefont {T.}~\bibnamefont {Taniguchi}}, \bibinfo {author}
  {\bibfnamefont {K.}~\bibnamefont {Watanabe}}, \bibinfo {author}
  {\bibfnamefont {K.}~\bibnamefont {Kadowaki}}, \bibinfo {author}
  {\bibfnamefont {R.~H.}\ \bibnamefont {Hadfield}}, \ and\ \bibinfo {author}
  {\bibfnamefont {D.}~\bibnamefont {Efetov}},\ }\bibfield  {title} {\enquote
  {\bibinfo {title} {Two-dimensional cuprate nanodetector with single telecom
  photon sensitivity at {$T=20\,$K}},}\ }\href {\doibase
  10.1038/s41598-020-66882-1} {\bibfield  {journal} {\bibinfo  {journal} {2D
  Mater.}\ }\textbf {\bibinfo {volume} {10}},\ \bibinfo {pages} {10256}
  (\bibinfo {year} {2023})}\BibitemShut {NoStop}%
\bibitem [{\citenamefont {Ziegler}\ \emph {et~al.}(2010)\citenamefont
  {Ziegler}, \citenamefont {Ziegler},\ and\ \citenamefont
  {Biersack}}]{Ziegler10}%
  \BibitemOpen
  \bibfield  {author} {\bibinfo {author} {\bibfnamefont {J.~F.}\ \bibnamefont
  {Ziegler}}, \bibinfo {author} {\bibfnamefont {M.~D.}\ \bibnamefont
  {Ziegler}}, \ and\ \bibinfo {author} {\bibfnamefont {J.~P}\ \bibnamefont
  {Biersack}},\ }\bibfield  {title} {\enquote {\bibinfo {title} {{SRIM} -- the
  stopping power and range of ions in matter},}\ }\href {\doibase
  10.1016/j.nimb.2010.02.091} {\bibfield  {journal} {\bibinfo  {journal}
  {Instrum. Methods Phys. Res., Sect. B}\ }\textbf {\bibinfo {volume} {268}},\
  \bibinfo {pages} {1818} (\bibinfo {year} {2010})}\BibitemShut {NoStop}%
\bibitem [{\citenamefont {Cui}\ \emph {et~al.}(1992)\citenamefont {Cui},
  \citenamefont {Xie},\ and\ \citenamefont {Li}}]{Cui92}%
  \BibitemOpen
  \bibfield  {author} {\bibinfo {author} {\bibfnamefont {F.~Z.}\ \bibnamefont
  {Cui}}, \bibinfo {author} {\bibfnamefont {J.}~\bibnamefont {Xie}}, \ and\
  \bibinfo {author} {\bibfnamefont {H.~D.}\ \bibnamefont {Li}},\ }\bibfield
  {title} {\enquote {\bibinfo {title} {Preferential radiation damage of the
  oxygen sublattice in {YBa$_2$Cu$_3$O$_{7}$}: A molecular-dynamics
  simulation},}\ }\href {\doibase 10.1103/PhysRevB.46.11182} {\bibfield
  {journal} {\bibinfo  {journal} {Phys. Rev. B}\ }\textbf {\bibinfo {volume}
  {46}},\ \bibinfo {pages} {11182--11185} (\bibinfo {year} {1992})}\BibitemShut
  {NoStop}%
\bibitem [{\citenamefont {Lang}\ \emph {et~al.}(2012)\citenamefont {Lang},
  \citenamefont {Marksteiner}, \citenamefont {Bodea}, \citenamefont {Siraj},
  \citenamefont {Pedarnig}, \citenamefont {Kolarova}, \citenamefont {Bauer},
  \citenamefont {Haselgr{\"u}bler}, \citenamefont {Hasenfuss}, \citenamefont
  {Beinik},\ and\ \citenamefont {Teichert}}]{Lang12}%
  \BibitemOpen
  \bibfield  {author} {\bibinfo {author} {\bibfnamefont {W.}~\bibnamefont
  {Lang}}, \bibinfo {author} {\bibfnamefont {M.}~\bibnamefont {Marksteiner}},
  \bibinfo {author} {\bibfnamefont {M.A.}\ \bibnamefont {Bodea}}, \bibinfo
  {author} {\bibfnamefont {K.}~\bibnamefont {Siraj}}, \bibinfo {author}
  {\bibfnamefont {J.D.}\ \bibnamefont {Pedarnig}}, \bibinfo {author}
  {\bibfnamefont {R.}~\bibnamefont {Kolarova}}, \bibinfo {author}
  {\bibfnamefont {P.}~\bibnamefont {Bauer}}, \bibinfo {author} {\bibfnamefont
  {K.}~\bibnamefont {Haselgr{\"u}bler}}, \bibinfo {author} {\bibfnamefont
  {C.}~\bibnamefont {Hasenfuss}}, \bibinfo {author} {\bibfnamefont
  {I.}~\bibnamefont {Beinik}}, \ and\ \bibinfo {author} {\bibfnamefont
  {C.}~\bibnamefont {Teichert}},\ }\bibfield  {title} {\enquote {\bibinfo
  {title} {Ion beam irradiation of cuprate high-temperature superconductors:
  Systematic modification of the electrical properties and fabrication of
  nanopatterns},}\ }\href {\doibase https://doi.org/10.1016/j.nimb.2011.01.087}
  {\bibfield  {journal} {\bibinfo  {journal} {Nucl. Instr. Meth. Phys. B}\
  }\textbf {\bibinfo {volume} {272}},\ \bibinfo {pages} {300--304} (\bibinfo
  {year} {2012})}\BibitemShut {NoStop}%
\bibitem [{\citenamefont {Mletschnig}\ and\ \citenamefont
  {Lang}(2019)}]{Mletschnig19}%
  \BibitemOpen
  \bibfield  {author} {\bibinfo {author} {\bibfnamefont {K.~L.}\ \bibnamefont
  {Mletschnig}}\ and\ \bibinfo {author} {\bibfnamefont {W.}~\bibnamefont
  {Lang}},\ }\bibfield  {title} {\enquote {\bibinfo {title} {Nano-patterning of
  cuprate superconductors by masked {He$^+$} ion irradiation: 3-dimensional
  profiles of the local critical temperature},}\ }\href {\doibase
  https://doi.org/10.1016/j.mee.2019.110982} {\bibfield  {journal} {\bibinfo
  {journal} {Microelectron.~Eng.}\ }\textbf {\bibinfo {volume} {215}},\
  \bibinfo {pages} {110982} (\bibinfo {year} {2019})}\BibitemShut {NoStop}%
\bibitem [{\citenamefont {Aichner}\ \emph {et~al.}(2019)\citenamefont
  {Aichner}, \citenamefont {M{\"u}ller}, \citenamefont {Karrer}, \citenamefont
  {Misko}, \citenamefont {Limberger}, \citenamefont {Mletschnig}, \citenamefont
  {Dosmailov}, \citenamefont {Pedarnig}, \citenamefont {Nori}, \citenamefont
  {Kleiner}, \citenamefont {Koelle},\ and\ \citenamefont {Lang}}]{Aichner19}%
  \BibitemOpen
  \bibfield  {author} {\bibinfo {author} {\bibfnamefont {Bernd}\ \bibnamefont
  {Aichner}}, \bibinfo {author} {\bibfnamefont {Benedikt}\ \bibnamefont
  {M{\"u}ller}}, \bibinfo {author} {\bibfnamefont {Max}\ \bibnamefont
  {Karrer}}, \bibinfo {author} {\bibfnamefont {Vyacheslav~R.}\ \bibnamefont
  {Misko}}, \bibinfo {author} {\bibfnamefont {Fabienne}\ \bibnamefont
  {Limberger}}, \bibinfo {author} {\bibfnamefont {Kristijan~L.}\ \bibnamefont
  {Mletschnig}}, \bibinfo {author} {\bibfnamefont {Meirzhan}\ \bibnamefont
  {Dosmailov}}, \bibinfo {author} {\bibfnamefont {Johannes~D.}\ \bibnamefont
  {Pedarnig}}, \bibinfo {author} {\bibfnamefont {Franco}\ \bibnamefont {Nori}},
  \bibinfo {author} {\bibfnamefont {Reinhold}\ \bibnamefont {Kleiner}},
  \bibinfo {author} {\bibfnamefont {Dieter}\ \bibnamefont {Koelle}}, \ and\
  \bibinfo {author} {\bibfnamefont {Wolfgang}\ \bibnamefont {Lang}},\
  }\bibfield  {title} {\enquote {\bibinfo {title} {Ultradense tailored vortex
  pinning arrays in superconducting {YBa$_2$Cu$_3$O$_{7-\delta}$} thin films
  created by focused he ion beam irradiation for fluxonics applications},}\
  }\href {\doibase 10.1021/acsanm.9b01006} {\bibfield  {journal} {\bibinfo
  {journal} {{ACS} Appl. Nano Mater.}\ }\textbf {\bibinfo {volume} {2}},\
  \bibinfo {pages} {5108--5115} (\bibinfo {year} {2019})}\BibitemShut {NoStop}%
\bibitem [{\citenamefont {Aichner}\ \emph {et~al.}(2020)\citenamefont
  {Aichner}, \citenamefont {Mletschnig}, \citenamefont {M\"{u}ller},
  \citenamefont {Karrer}, \citenamefont {Dosmailov}, \citenamefont {Pedarnig},
  \citenamefont {Kleiner}, \citenamefont {Koelle},\ and\ \citenamefont
  {Lang}}]{Aichner20}%
  \BibitemOpen
  \bibfield  {author} {\bibinfo {author} {\bibfnamefont {B.}~\bibnamefont
  {Aichner}}, \bibinfo {author} {\bibfnamefont {K.~L.}\ \bibnamefont
  {Mletschnig}}, \bibinfo {author} {\bibfnamefont {B.}~\bibnamefont
  {M\"{u}ller}}, \bibinfo {author} {\bibfnamefont {M.}~\bibnamefont {Karrer}},
  \bibinfo {author} {\bibfnamefont {M.}~\bibnamefont {Dosmailov}}, \bibinfo
  {author} {\bibfnamefont {J.~D.}\ \bibnamefont {Pedarnig}}, \bibinfo {author}
  {\bibfnamefont {R.}~\bibnamefont {Kleiner}}, \bibinfo {author} {\bibfnamefont
  {D.}~\bibnamefont {Koelle}}, \ and\ \bibinfo {author} {\bibfnamefont
  {W.}~\bibnamefont {Lang}},\ }\bibfield  {title} {\enquote {\bibinfo {title}
  {Angular magnetic-field dependence of vortex matching in pinning lattices
  fabricated by focused or masked helium ion beam irradiation of
  superconducting {YBa$_2$Cu$_3$O$_{7-\delta}$} thin films},}\ }\href {\doibase
  10.1063/10.0000863} {\bibfield  {journal} {\bibinfo  {journal} {Low Temp.
  Phys.}\ }\textbf {\bibinfo {volume} {46}},\ \bibinfo {pages} {331--337}
  (\bibinfo {year} {2020})}\BibitemShut {NoStop}%
\bibitem [{\citenamefont {Cho}\ \emph {et~al.}(2016)\citenamefont {Cho},
  \citenamefont {Kouperine}, \citenamefont {Zhuo}, \citenamefont {Dynes},\ and\
  \citenamefont {Cybart}}]{Cho16}%
  \BibitemOpen
  \bibfield  {author} {\bibinfo {author} {\bibfnamefont {E.~Y.}\ \bibnamefont
  {Cho}}, \bibinfo {author} {\bibfnamefont {K.}~\bibnamefont {Kouperine}},
  \bibinfo {author} {\bibfnamefont {Y.}~\bibnamefont {Zhuo}}, \bibinfo {author}
  {\bibfnamefont {R.~C.}\ \bibnamefont {Dynes}}, \ and\ \bibinfo {author}
  {\bibfnamefont {S.~A.}\ \bibnamefont {Cybart}},\ }\bibfield  {title}
  {\enquote {\bibinfo {title} {The effects of annealing a 2-dimensional array
  of ion-irradiated {J}osephson junctions},}\ }\href {\doibase
  10.1088/0953-2048/29/9/094004} {\bibfield  {journal} {\bibinfo  {journal}
  {Supercond. Sci. Technol}\ }\textbf {\bibinfo {volume} {29}},\ \bibinfo
  {pages} {094004} (\bibinfo {year} {2016})}\BibitemShut {NoStop}%
\bibitem [{\citenamefont {Miller}\ \emph {et~al.}(2022)\citenamefont {Miller},
  \citenamefont {Lemon}, \citenamefont {Choffel}, \citenamefont {Rich},
  \citenamefont {Harvel},\ and\ \citenamefont {Johnson}}]{Miller22}%
  \BibitemOpen
  \bibfield  {author} {\bibinfo {author} {\bibfnamefont {A.~M.}\ \bibnamefont
  {Miller}}, \bibinfo {author} {\bibfnamefont {M.}~\bibnamefont {Lemon}},
  \bibinfo {author} {\bibfnamefont {M.~A.}\ \bibnamefont {Choffel}}, \bibinfo
  {author} {\bibfnamefont {S.~R.}\ \bibnamefont {Rich}}, \bibinfo {author}
  {\bibfnamefont {F.}~\bibnamefont {Harvel}}, \ and\ \bibinfo {author}
  {\bibfnamefont {D.~C.}\ \bibnamefont {Johnson}},\ }\bibfield  {title}
  {\enquote {\bibinfo {title} {Extracting information from {X}-ray diffraction
  patterns containing {L}aue oscillations},}\ }\href {\doibase
  10.1515/znb-2022-0020} {\bibfield  {journal} {\bibinfo  {journal} {Z.
  Naturforsch. B}\ }\textbf {\bibinfo {volume} {77}},\ \bibinfo {pages}
  {313--322} (\bibinfo {year} {2022})}\BibitemShut {NoStop}%
\bibitem [{\citenamefont {Hooker}\ \emph {et~al.}(1993)\citenamefont {Hooker},
  \citenamefont {Wise}, \citenamefont {Carlberg}, \citenamefont {Stephens},
  \citenamefont {Simchick},\ and\ \citenamefont {Farjami}}]{Hooker93}%
  \BibitemOpen
  \bibfield  {author} {\bibinfo {author} {\bibfnamefont {M.~W.}\ \bibnamefont
  {Hooker}}, \bibinfo {author} {\bibfnamefont {S.~A.}\ \bibnamefont {Wise}},
  \bibinfo {author} {\bibfnamefont {I.~A.}\ \bibnamefont {Carlberg}}, \bibinfo
  {author} {\bibfnamefont {R.~M.}\ \bibnamefont {Stephens}}, \bibinfo {author}
  {\bibfnamefont {R.~T.}\ \bibnamefont {Simchick}}, \ and\ \bibinfo {author}
  {\bibfnamefont {A.}~\bibnamefont {Farjami}},\ }\href@noop {} {\emph {\bibinfo
  {title} {Room temperature degradation of {YBa$_2$Cu$_3$O$_{7-x}$}
  superconductors in varying relative humidity environments}}},\ \bibinfo
  {type} {Technical Paper}\ \bibinfo {number} {3368}\ (\bibinfo  {institution}
  {NASA},\ \bibinfo {year} {1993})\BibitemShut {NoStop}%
\bibitem [{\citenamefont {Mogro-Campero}\ \emph {et~al.}(1995)\citenamefont
  {Mogro-Campero}, \citenamefont {Paik},\ and\ \citenamefont
  {Turner}}]{Mogro-Campero95}%
  \BibitemOpen
  \bibfield  {author} {\bibinfo {author} {\bibfnamefont {A.}~\bibnamefont
  {Mogro-Campero}}, \bibinfo {author} {\bibfnamefont {K.~W.}\ \bibnamefont
  {Paik}}, \ and\ \bibinfo {author} {\bibfnamefont {L.~G.}\ \bibnamefont
  {Turner}},\ }\bibfield  {title} {\enquote {\bibinfo {title} {Degradation of
  thin films of {YBa$_2$Cu$_3$O$_7$} by annealing in air and in vacuum},}\
  }\href {\doibase 10.1007/BF00732247} {\bibfield  {journal} {\bibinfo
  {journal} {J. Supercond.}\ }\textbf {\bibinfo {volume} {8}},\ \bibinfo
  {pages} {95--98} (\bibinfo {year} {1995})}\BibitemShut {NoStop}%
\bibitem [{\citenamefont {Chesca}\ \emph {et~al.}(2004)\citenamefont {Chesca},
  \citenamefont {Kleiner},\ and\ \citenamefont {Koelle}}]{Chesca-SHB-2}%
  \BibitemOpen
  \bibfield  {author} {\bibinfo {author} {\bibfnamefont {B.}~\bibnamefont
  {Chesca}}, \bibinfo {author} {\bibfnamefont {R.}~\bibnamefont {Kleiner}}, \
  and\ \bibinfo {author} {\bibfnamefont {D.}~\bibnamefont {Koelle}},\ }\enquote
  {\bibinfo {title} {{SQUID} {T}heory},}\ in\ \href@noop {} {\emph {\bibinfo
  {booktitle} {The {SQUD} {H}andbook}}},\ Vol.\ \bibinfo {volume} {1:
  Fundamentals and Technology of SQUIDs and SQUID systems},\ \bibinfo {editor}
  {edited by\ \bibinfo {editor} {\bibfnamefont {John}\ \bibnamefont {Clarke}}\
  and\ \bibinfo {editor} {\bibfnamefont {Alex~I.}\ \bibnamefont {Braginski}}}\
  (\bibinfo  {publisher} {Wiley-VCH},\ \bibinfo {address} {Weinheim},\ \bibinfo
  {year} {2004})\ Chap.~\bibinfo {chapter} {2}, pp.\ \bibinfo {pages}
  {29--92}\BibitemShut {NoStop}%
\bibitem [{not()}]{note1}%
  \BibitemOpen
  \href@noop {} {}\bibinfo {note} {For the two highest dose values $D=800$ and
  $1000\units{i/nm}$, the $\ls{j}{c}(t)$ data shown in fig.~\ref{fig:time-jc}
  were at $t=288\units{days}$ still far from saturation. Therefore, the values
  of $\ls{j}{c,\infty}$ determined from the fit have huge uncertainties, i.e.,
  one can fit the experimental data quite well with many combinations of
  $\ls{j}{c,\infty}$ and $\tau$. Therefore, for the two highest doses the
  $\ls{j}{c,\infty}$ values (open symbols in fig.~\ref{fig:fitparams}(a)) have
  been fixed by hand so that they lay exactly on the line shown in
  fig.~\ref{fig:fitparams}(a) (open circles). After fixing $\ls{j}{c,\infty}$
  for the highest two D-values, the parameters $b$ and $\tau$ were calculated
  from the fits of $\f{\ls{j}{c}}{t}$.}\BibitemShut {Stop}%
\end{thebibliography}%

\end{document}